%% file: FWD-10-008_temp.tex
\pdfoutput=1

\documentclass[11pt,twoside,a4paper,cmspaper,final,collab]{cms-tdr}
\def\svnVersion{79929:80115}\def\cmsCernNo{2011-141}\def\cmsCernDate{\today}\def\cmsMessage{Submitted to the European Physical Journal C}

\begin{document}\cmsNoteHeader{FWD-10-008}

\hyphenation{had-ron-i-za-tion}
\hyphenation{cal-or-i-me-ter}
\hyphenation{de-vices}
\def\gtap{\raisebox{-.4ex}{\rlap{$\sim$}} \raisebox{.4ex}{$>$}}
\newcommand {\pom} {I\!\!P}
\newcommand {\reg} {I\!\!R}
\providecommand{\POMPYT}{\textsc{pompyt}\xspace}
\newcommand{\PYTHIAsix}{\textsc{pythia6}\xspace}
\newcommand{\PYTHIAeight}{\textsc{pythia6}\xspace}

\RCS$Revision: 80115 $
\RCS$HeadURL: svn+ssh://alverson@svn.cern.ch/reps/tdr2/papers/FWD-10-008/trunk/FWD-10-008.tex $
\RCS$Id: FWD-10-008.tex 80115 2011-10-02 13:01:52Z alverson $

\cmsNoteHeader{FWD-10-008}
\title{Forward Energy Flow, Central Charged-Particle Multiplicities, and Pseudorapidity Gaps
 in $\PW$ and $\cPZ$ Boson Events from $\Pp\Pp$ Collisions at $\sqrt{s}= 7$~TeV}
\ifthenelse{\boolean{cms@external}}{\titlerunning{Forward Energy Flow, \ldots}}{\relax}

\date{\today}

\abstract{
A study of forward energy flow and central charged-particle
multiplicity in events with $\PW$ and $\cPZ$ bosons decaying into leptons
is presented. The analysis uses a sample of 7 TeV $\Pp\Pp$ collisions,
corresponding to an integrated luminosity of 36 pb$^{-1}$, recorded by the CMS experiment at the LHC.
The observed forward energy depositions, their correlations, and the central
charged-particle multiplicities are not well described
by the available non-diffractive soft-hadron production models.
A study of about 300 events with no significant energy deposited in one of the forward calorimeters,
corresponding to a pseudorapidity gap of at least 1.9 units, is also presented.
An indication for a diffractive component in these events comes from the
observation that the majority of the charged leptons from the $\PW(\cPZ)$ decays are found in the hemisphere opposite to the gap. When fitting the signed lepton pseudorapidity distribution
of these
events
with predicted distributions from
an admixture of diffractive (\POMPYT) and non-diffractive (\PYTHIA) Monte Carlo simulations,
the diffractive component is determined to be (50.0~$\pm$~9.3~(stat.)~$\pm$~5.2 (syst.))\%.
}

\hypersetup{%
pdfauthor={CMS Collaboration},%
pdftitle={Forward Energy Flow, Central Charged-Particle Multiplicities, and Pseudorapidity Gaps
 in W and Z Boson Events from pp Collisions at 7 TeV}, %
pdfsubject={Electroweak Physics, Forward Physics},%
pdfkeywords={CMS, physics, vector bosons, diffraction, rapidity gaps}}

\maketitle %maketitle comes after all the front information has been supplied

\section{Introduction}

At high energies, proton-proton reactions are generally described in terms of a two-component outgoing system, $\Pp\Pp \rightarrow XY$, where $X$ is a state originating from a perturbative parton-parton interaction and $Y$ consists of proton remnants carrying a large fraction of the total energy. The system $Y$ represents the ``underlying event" and consists of mostly low-transverse-momentum hadrons originating from parton showers, and non-perturbative multi-parton interactions.
The latter are largely independent of the hard interaction, increase the particle multiplicity, and lead to correlations between the energy flows in the central and forward regions.

The currently available models of multi-parton interactions
have mainly been tuned to
minimum-bias data and to final states including jets with large transverse momentum ($p_{T}$),
using the observed central charged-particle multiplicities and the transverse momentum spectra
of hadrons that are not associated with the hard jets. Models for a detailed simulation of
multi-parton interactions are under rapid development and new features, including
diffractive components in the energy flow, are being extensively investigated.
A recent detailed CMS study of  the underlying event structure at central rapidities is given in
\cite{QCD-10-10}.

The analysis of the underlying event structure and energy flow correlations
in hard processes with colorless final states, such as
$\Pp\Pp \rightarrow \PW(\cPZ) X \rightarrow \ell \nu (\ell \ell) X$, can provide insights into
unexplored aspects of multi-parton interactions.
In particular, processes with colorless final states
allow a straightforward separation of the hard interaction and the underlying event.
Correlations between the charged-particle multiplicity in the central rapidity range and energy depositions at large rapidities can give additional information about multi-parton interactions.

Furthermore, a fraction of these proton-proton interactions is expected to arise from
single-diffractive (SD) reactions, where one of the colliding protons emerges intact from the interaction, having lost only a few percent of its energy. Such SD events may be ascribed to the exchange of vacuum quantum numbers (often called Pomeron exchange), which leads to the absence of hadron production over a wide region of rapidity adjacent to the outgoing proton direction.
Experimentally, these large rapidity gaps will appear as regions of pseudorapidity, devoid of detected particles.

Soft-diffractive events can be described in the framework of Regge theory (see, e.g.,~\cite{Collins:1977jy}).
Hard-diffractive events, with the production of jets, heavy flavors, or $\PW/\cPZ$ bosons, have been observed at the SPS, HERA, and the Tevatron~\cite{Brandt:1997gi, :2009qja,
H1:2007,Aktas:2006hy,
Abe:1997jp,Abazov:2003ti}. For electron-proton hard-diffractive interactions, a factorization theorem has been proven~\cite{Collins:1997sr}, allowing the introduction of diffractive parton distribution functions (dPDFs). In hadron-hadron diffractive interactions, factorization is however broken by soft multi-parton
interactions~\cite{Bjorken:1992er,Kaidalov:2001iz}, which fill the large rapidity gap and
reduce the observed yields of hard-diffractive events. Such interactions are not yet simulated in the currently available Monte Carlo (MC) generators, and the reduction in the large rapidity gap hard-diffraction cross section is quantified by a suppression factor, the so-called rapidity gap survival probability.
At the Tevatron, the observed
hard-diffractive yields relative to the corresponding inclusive processes are approximately 1\%.
A recent study by CDF~\cite{Aaltonen:2010qe} indicates that the fractions of $\PW$ and $\cPZ$ bosons produced diffractively are (1.00 $\pm$ 0.11)\% and (0.88 $\pm$ 0.22)\%, respectively.

\begin{figure*}[!Hhtb]
\centering

\subfigure[]{\label{fig:diag-y}\includegraphics[width=0.26\textwidth]{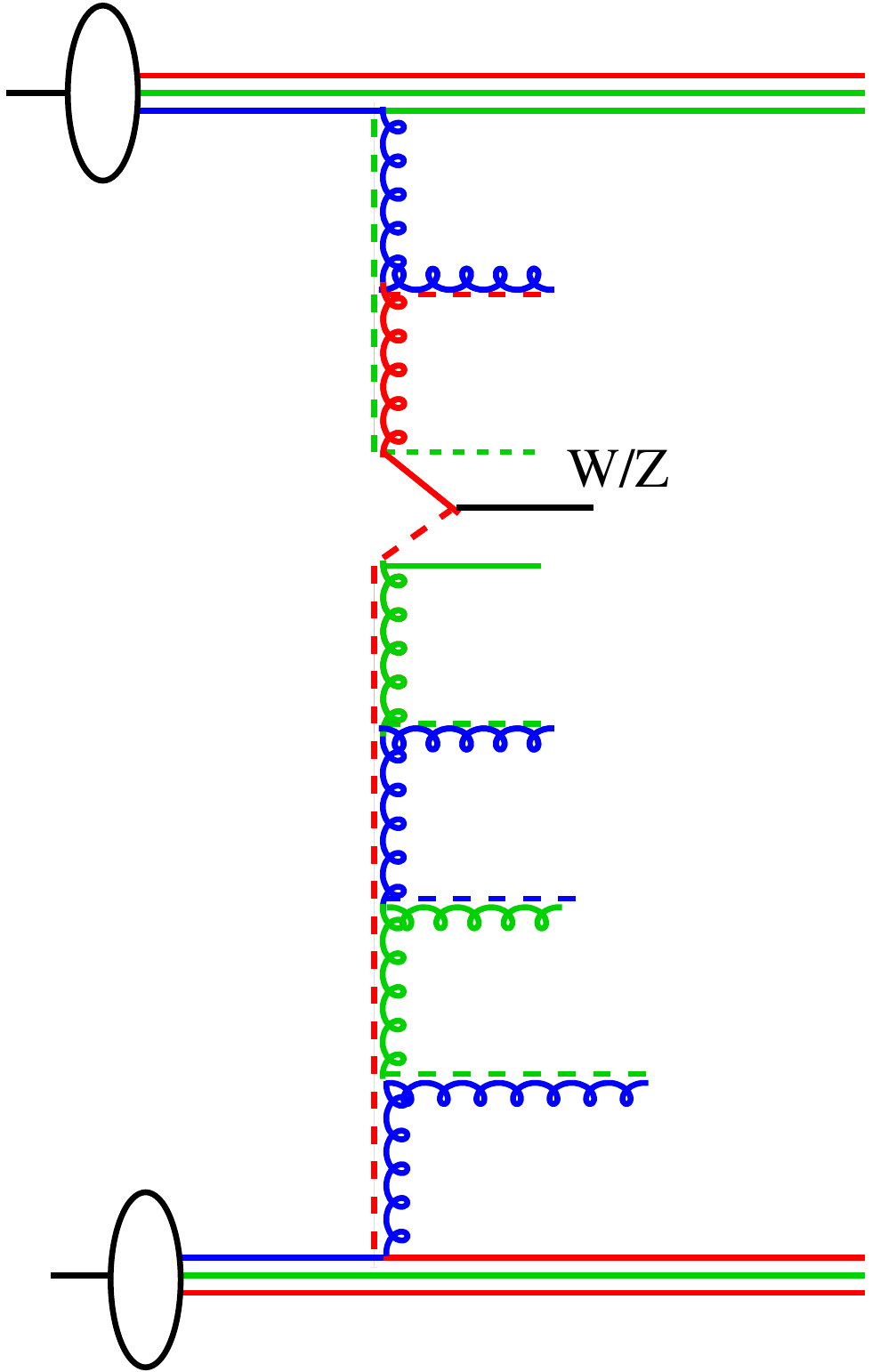}}
\subfigure[]{\label{fig:diag-a}\includegraphics[width=0.3\textwidth]{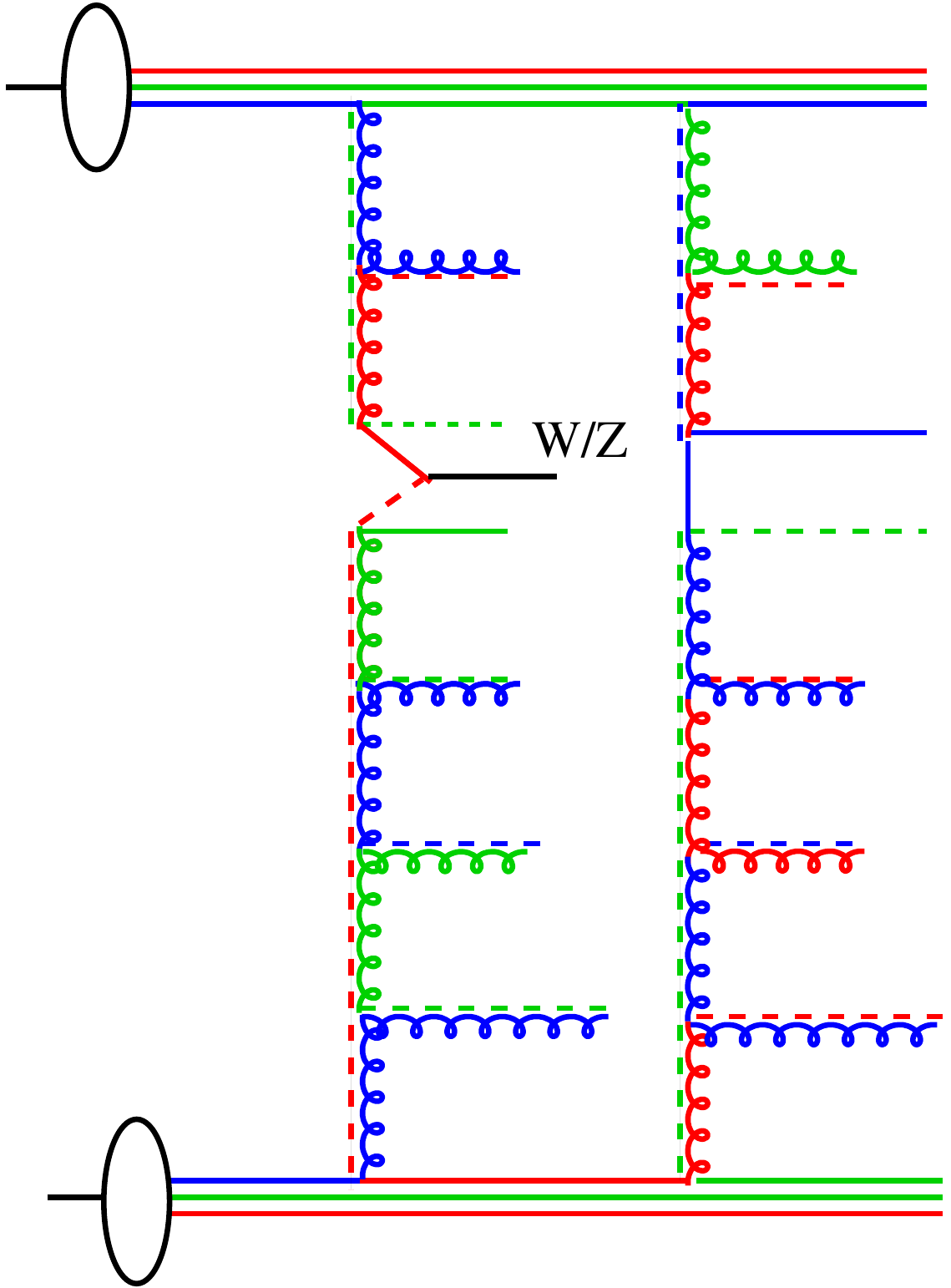}}
\subfigure[]{\label{fig:diag-c}\includegraphics[width=0.3\textwidth]{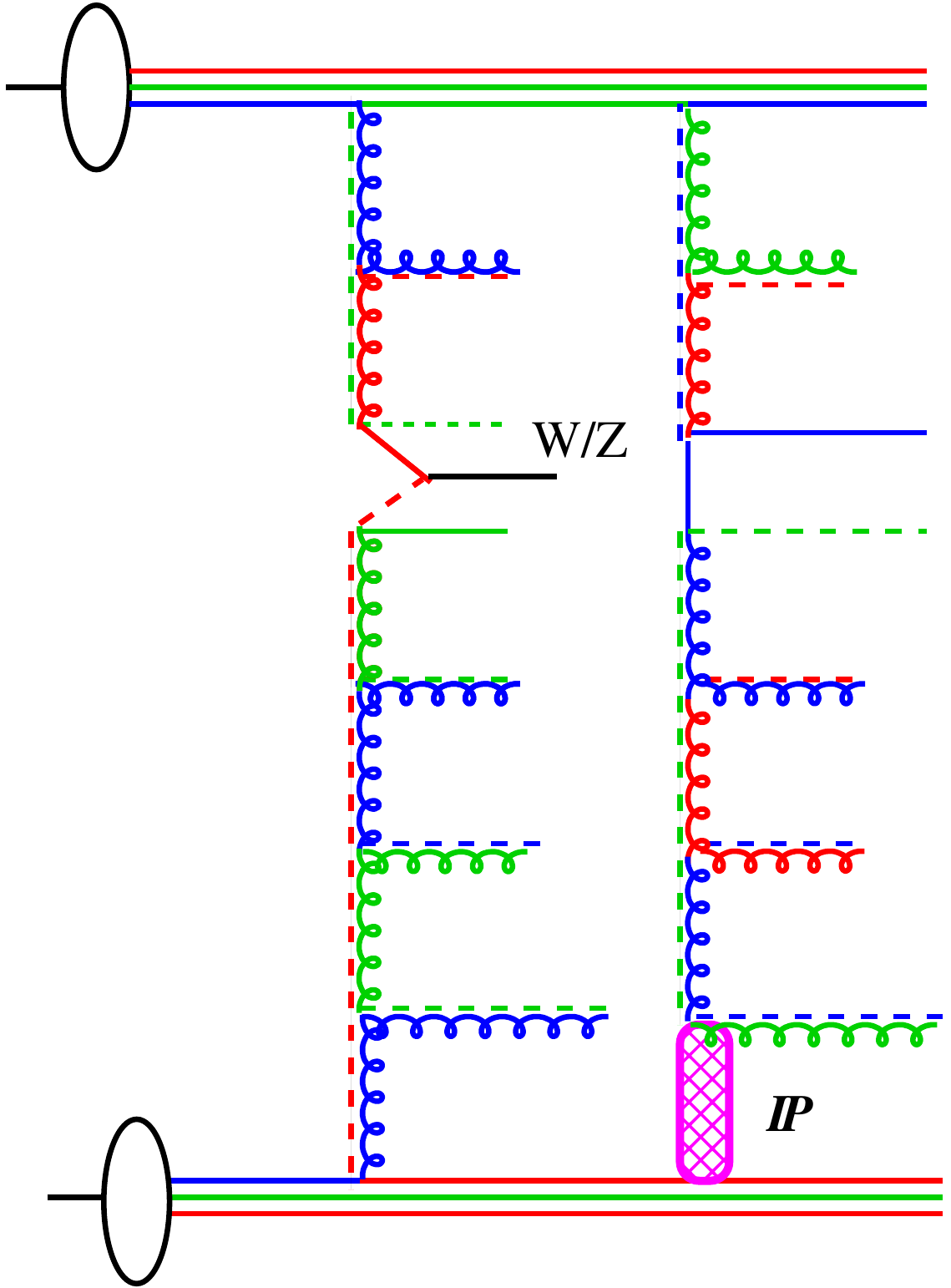}} \\
\subfigure[]{\label{fig:diag-z}\includegraphics[width=0.3\textwidth]{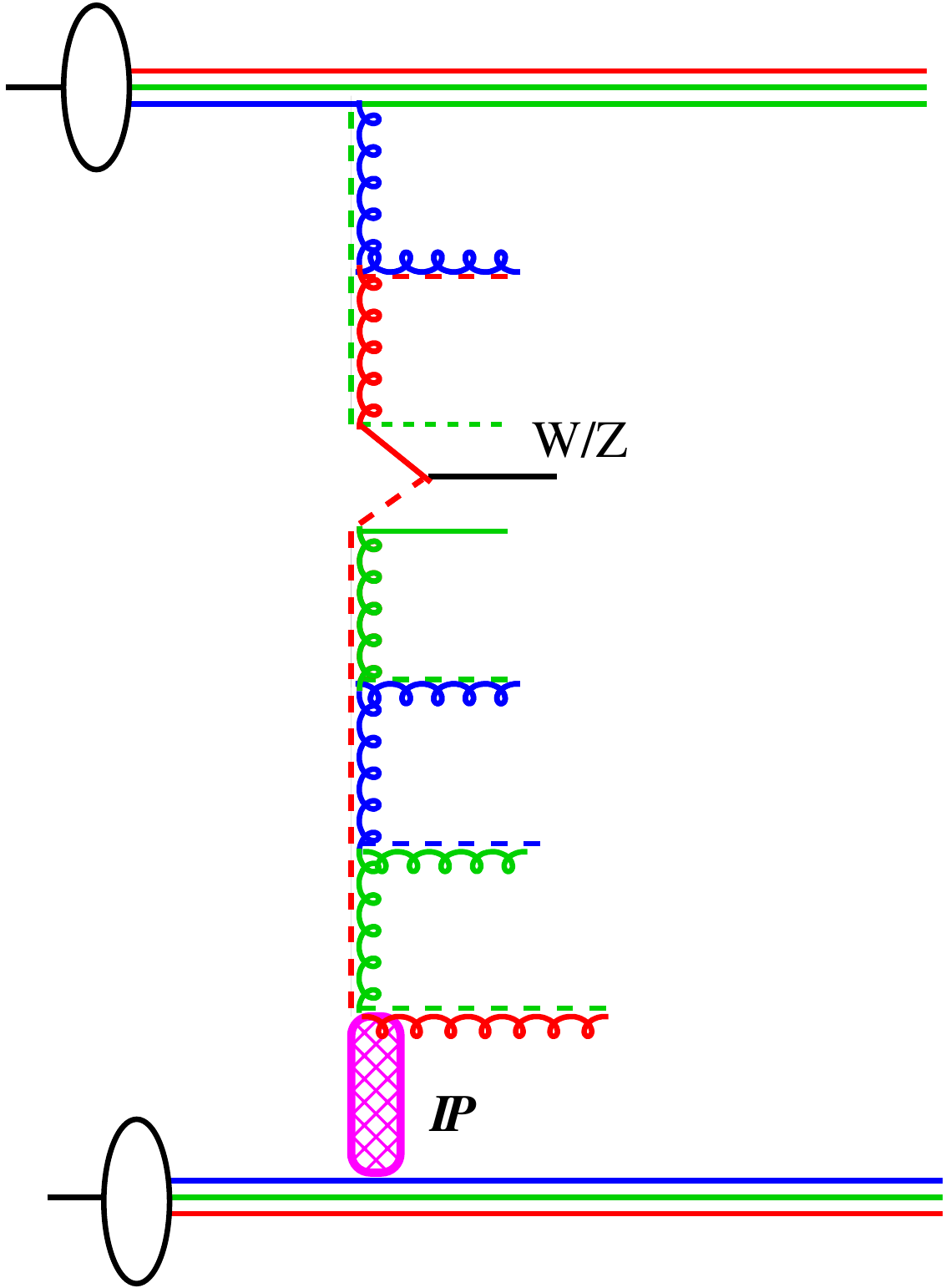}}
\subfigure[]{\label{fig:diag-b}\includegraphics[width=0.3\textwidth]{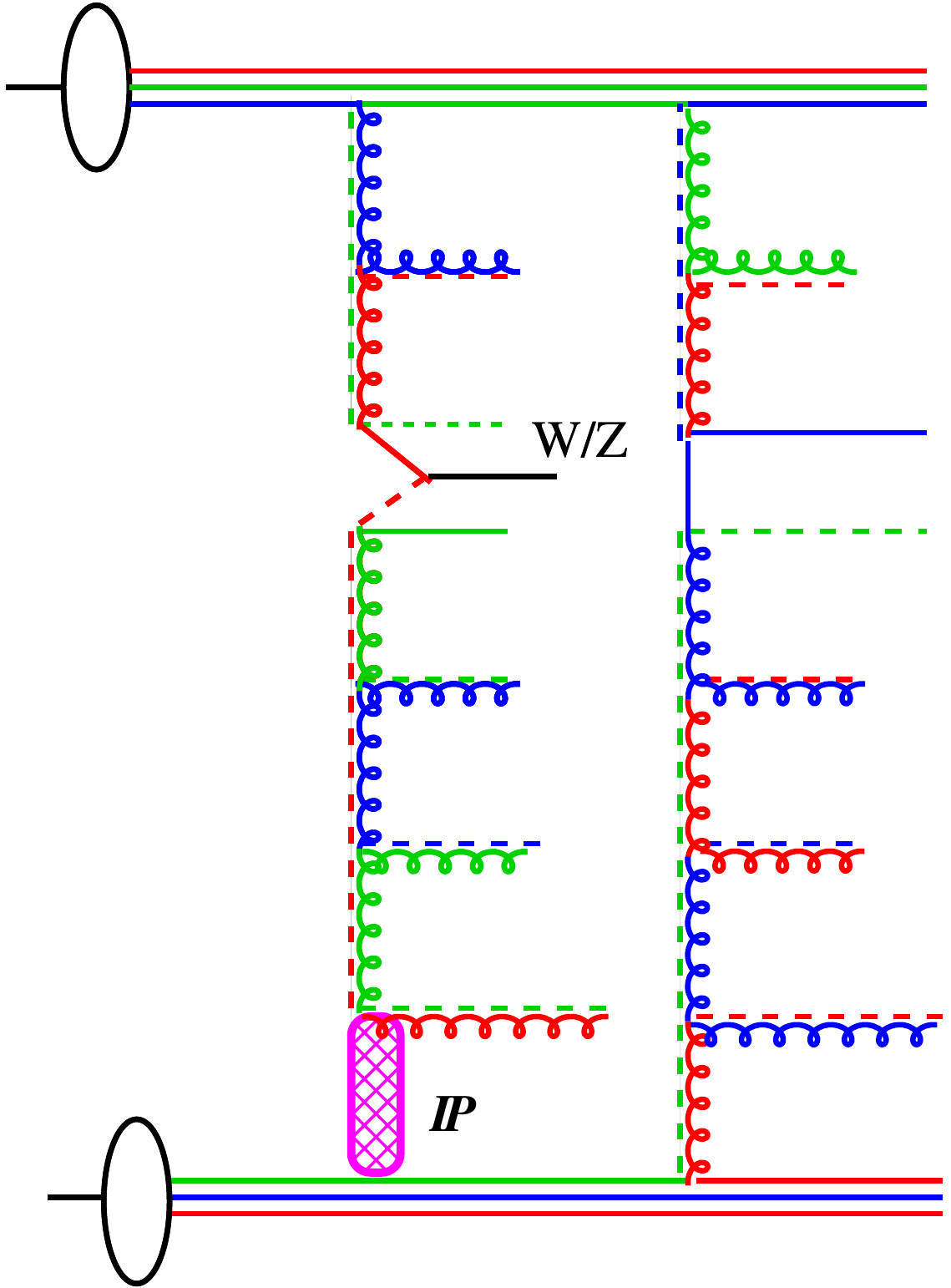}}
\subfigure[]{\label{fig:diag-e}\includegraphics[width=0.3\textwidth]{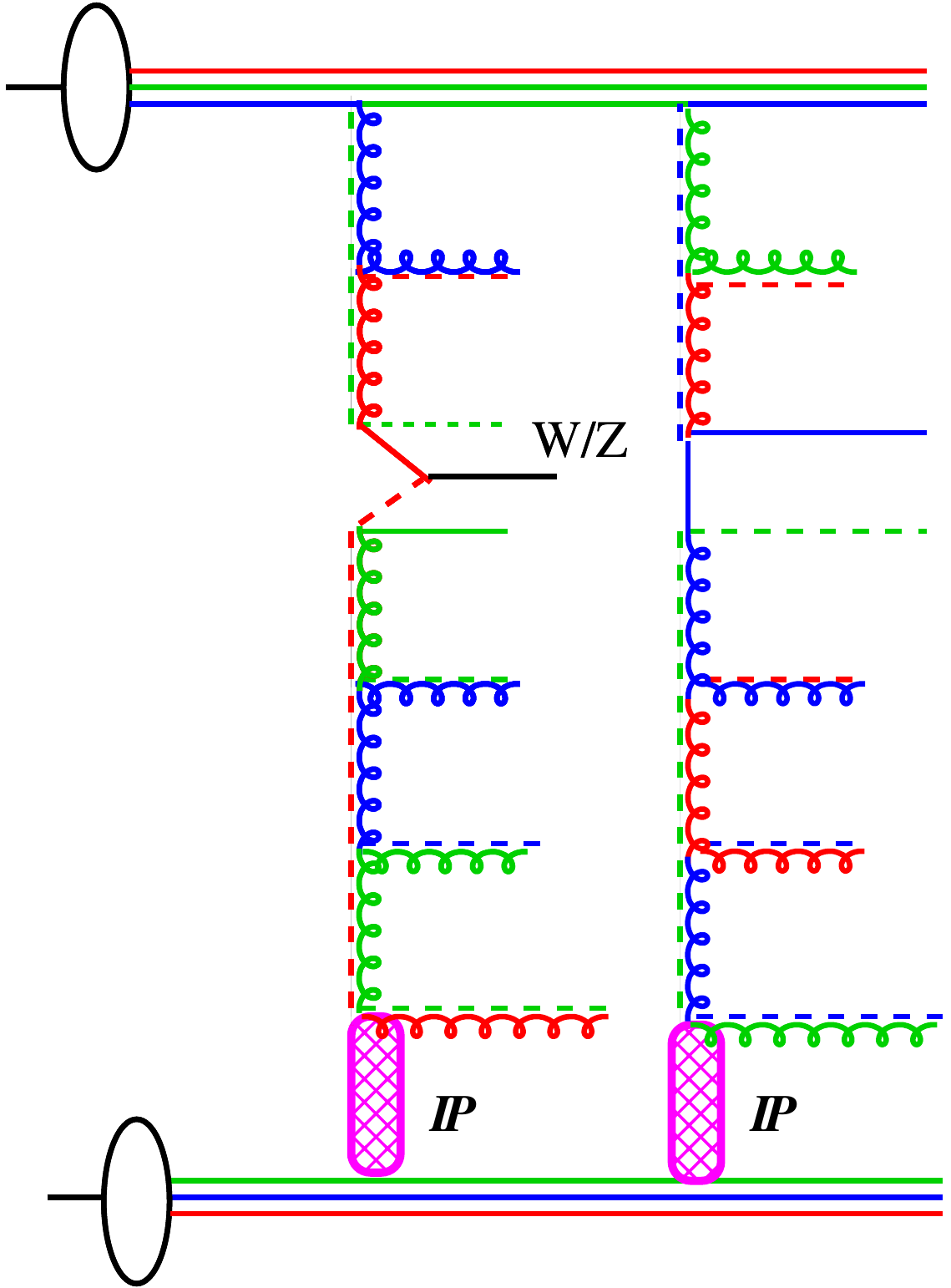}}

\caption{Sketches of the process $\Pp\Pp \to \PW(\cPZ)X$, where a $\PW(\cPZ)$
boson is produced in the hard interaction, combined with contributions
from multi-parton interactions, with and without a diffractive component. The colored curly and straight lines
are for gluons and quarks and the straight black lines are for $\PW(\cPZ)$.
The symbol $\pom$ indicates the exchange of a state with the quantum numbers of the vacuum (Pomeron).  (a) shows the standard hard interaction; (b) the same process with additional
multi-parton interactions; (c) the hard process accompanied by multi-parton interactions containing
a diffractive component; (d) the hard-diffractive production of a $\PW(\cPZ)$ boson;
(e) the hard-diffractive $\PW(\cPZ)$ production with multi-parton interactions, and (f)
the hard-diffractive $\PW(\cPZ)$ production with multi-parton interactions containing
a diffractive component. In the latter case, the diffractive component of the multi-parton interaction does not necessarily couple to the same proton
as in the hard process (not shown here).
}
\label{Fig1-diagram}
\end{figure*}

The standard picture of $\PW$ or $\cPZ$ boson production via hard parton-parton scattering
is shown in Fig. \ref{fig:diag-y} and combined with a multi-parton interaction in Fig. \ref{fig:diag-a}.  According to this
picture, large rapidity gap events can only arise from multiplicity fluctuations.  Figure \ref{fig:diag-c} shows standard $\PW(\cPZ)$ production accompanied by multi-parton interactions with a diffractive component.
Hard-diffractive production (Fig. \ref{fig:diag-z}) leads to a large rapidity gap. Such contributions would result in almost unchanged central charged-particle multiplicity distributions with respect
to Fig. \ref{fig:diag-a}.
However, a larger fraction of events with a relatively small energy deposition in the forward regions, and a smaller correlation of the energy flow in the central
and forward regions, could be expected.
The diffractive production mechanism with multi-parton interactions is shown in
Fig. \ref{fig:diag-b}. In this case, large rapidity gap events only survive if the multi-parton component is
small. Finally, Fig. \ref{fig:diag-e} indicates a possible combination of  $\PW(\cPZ)$
production mechanisms with a diffractive component in both the hard process and the multi-parton interaction.

This paper presents an analysis of the underlying event structure and of events with a
pseudorapidity gap of more than 1.9 units (in the following, referred to as LRG events for simplicity)
in the processes $\Pp\Pp \rightarrow \PW X$ and $\Pp\Pp \rightarrow \cPZ X$, in 7 TeV $\Pp\Pp$ collisions at the
Large Hadron Collider (LHC), and
recorded in 2010 with the Compact Muon Solenoid (CMS) detector. A data sample
corresponding to an integrated luminosity of 36 pb$^{-1}$ has been analyzed.
Final states of the type $\PW X$ and $\cPZ X$ are identified by the detection of electrons
and muons with large transverse momentum $p_{T}$.
The underlying event structure is analyzed in terms of
(i) the charged-particle multiplicity at central rapidities, (ii) the forward energy flow,
and (iii) the correlations between these observables.
The observed LRG events are expected to be sensitive to a
diffractive production component.

The paper is structured as follows:
after giving a short description of the models for inclusive and diffractive
vector-boson production in $\Pp\Pp$ collisions (Section 2) and the CMS detector (Section 3),
the selection of $\PW$ and $\cPZ$ bosons with leptonic decays, the rejection of pileup events, and the energy measurement in the forward calorimeters are described in Section 4.
Section 5 summarizes the analysis of the underlying event structure in
$\PW X$ and $\cPZ X$ events in terms of the observed charged-particle multiplicity, the energy depositions in the forward
calorimeters, and their correlations. The analysis of events with a LRG signature is discussed in Section 6.

\section{Modeling of the underlying event structure in \texorpdfstring{$\mathbf{W}$ and $\mathbf{Z}$}{W and Z} events}

The simulation of non-diffractive processes, including multiple overlaid events within the same
bunch crossing (``pileup"),
was performed using the {\PYTHIA 6.420} and {\PYTHIA 8.145} event generators
\cite{Sjostrand:2006za,pythia8} with
different tunes~\cite{DESY-PROC-2009-06} for the underlying event structure and the multi-parton interactions. Several tunes were used for the
simulation of the underlying event structure in $\PW$ and $\cPZ$ events. In particular, the tunes developed
before data from the LHC could be used
were {\PYTHIAsix} D6T \cite{Field:2010su}, Pro-Q20~\cite{Buckley:2009bj}, Pro-PT0~\cite{Buckley:2009bj}, and P0 \cite{Skands:2010ak}. The newer tunes
{\PYTHIAsix} Z2 \cite{Field:2010bc} and the {\PYTHIAeight} 2C \cite{Corke:2010yf} include already
some information from the LHC data.
It is relevant for the discussion in Section 5 that the older D6T and Pro-Q20 tunes are associated with virtuality-ordered showers, while the newer ones, P0, Pro-PT0, Z2, and 2C, are associated with $p_{T}$-ordered showers.
As shown in the recent CMS measurement of the underlying event structure
Ref.~\cite{QCD-10-10}, the Z2 tune provides a reasonably good description of the data in
minimum-bias events.

Using the average instantaneous luminosity for each running period (discussed below),
an average number of
soft pileup minimum-bias MC events were superimposed on the simulated hard-interaction events.
As discussed in detail in Section 5, the predicted pileup contribution to the forward energy flow
from the simulation
was found to be in good agreement with the one from a dedicated unbiased data sample.

The $\PW$ and $\cPZ$ production cross sections were calculated using leading-order (LO) matrix elements
for the process $ \cPq  \cPaq \rightarrow \PW (\cPZ)$, convolved with the CTEQ parton
distribution functions (PDFs)~\cite{Nadolsky:2008zw}.
Effects from higher-order QCD corrections were approximated with parton showers from the
initial and final-state partons.

Diffractive $\PW$ and $\cPZ$ production was simulated with the \POMPYT 2.6.1 \cite{Bruni:1993ym,Bruni:1993is} event generator. The hard processes responsible for the production of $\PW$ and $\cPZ$ bosons were identical to those in non-diffractive models.
For the simulation of diffractive processes, the dPDFs (fit B) measured by the H1 experiment at HERA
were used~\cite{Aktas:2006hy, Aktas:2007bv}.
This generator does not simulate multi-parton interactions or the ensuing rapidity gap survival probability.

\section{Experimental apparatus}

A detailed description of the CMS experiment can
be found elsewhere~\cite{:2008zzk}.
The coordinate system has the origin at the nominal interaction point. The $\cPZ$-axis is parallel to the anticlockwise beam direction; it defines the polar angle $\theta$ and the pseudorapidity
$ \eta  = - \ln(\tan(\theta/2))$. The azimuthal angle $\phi$ is measured in the plane transverse to the beam, from the direction pointing to the centre of the LHC ring toward the upward direction.

The central feature of the CMS
apparatus is a superconducting solenoid, of 6~m internal diameter. Within
the field volume are the silicon pixel and strip tracker, the crystal
electromagnetic calorimeter (ECAL), and the brass scintillator hadronic
calorimeter (HCAL).  Muons are measured in gaseous detectors embedded in
the steel return yoke. The ECAL, HCAL, and muon detectors are composed of barrel
and endcap sections.
The calorimeter cells are grouped in projective towers, of
granularity $\Delta \eta \times \Delta \phi = 0.087\times0.087$ radians at central
rapidities. For forward rapidities the HCAL towers have a granularity of 0.174 radians in $\phi$ and
are increasing as a function of $\eta$.
Besides the
barrel and endcap detectors, CMS has extensive forward calorimetry. The
forward hadronic calorimeter (HF) consists of steel absorbers
and embedded radiation-hard quartz fibers, which provide fast collection
of Cherenkov light.  The maximum pseudorapidity coverage is $2.9<|\eta|<5.2$,
but because of small differences in the passive material at large values of
$|\eta|$ between the real detector and the
simulation, a fiducial coverage of $3.0<|\eta|<4.9$ is used.

\section{Event selection procedure}
\label{sec:eventselection}

\subsection{Selection of \texorpdfstring{$\mathbf{W}$ and $\mathbf{Z}$}{W and Z} events with decays to electrons and muons}

The identification of $\PW$ and $\cPZ$ bosons is based on the presence of isolated electrons and muons with high transverse momentum.
The selection criteria are given below. A more detailed description of the lepton selection,
the efficiencies, and the associated systematic uncertainties is given in \cite{Khachatryan:2010xn}.

Events are selected online by requiring a high-transverse-momentum electron or muon
with thresholds depending on the run period and varying between 10 and 17 GeV for electrons and between 9 and 15 GeV for muons.
Thus, a small fraction of $\PW \to \tau \nu$ or $\cPZ \to \tau \tau$ events
with leptonic decays of the $\tau$ is also included. The trigger efficiency for signal events with the selection conditions defined below
is above 99\% \cite{Khachatryan:2010xn}.

The offline selection of electrons is based on the matching of a reconstructed
high-transverse-momentum track candidate with energy depositions in the barrel and endcap calorimeters; shower shape requirements are applied to these clusters. Electron candidates are required to have
$|\eta| < 2.5$ and
transverse momenta larger than 25 GeV. Electrons from photon conversions are rejected.
In order to suppress background from jets, the electrons are required to be isolated using a
cone in $\eta-\phi$ space around the electron direction
with $\Delta R < 0.3$, where  $\Delta R = \sqrt{\Delta \eta^{2} + \Delta \phi^{2}}$ and
$\Delta \eta$ and $\Delta \phi$ are the pseudorapidity and azimuthal angle difference between the electron
and jet directions.
The energy in the cone is calculated from the scalar sum of the transverse energies in the tracker and both calorimeters, and is required to be smaller than 10\% of the lepton $p_{T}$.
More details are given in Ref. \cite{Khachatryan:2010xn}.

Two algorithms are used to identify muons. One is based on the matching of a reconstructed
high-transverse-momentum silicon tracker candidate with a track candidate found in the muon system.
The second is based on a global fit to tracker and muon system hits. Muons in this analysis have to pass both selection algorithms. The reconstructed muon candidates are required to be at $|\eta| < 2.5$
and to have transverse momentum larger than 25 GeV.
The isolation criteria are similar to the ones used for electrons, and details about the muon
identification are given in Ref. \cite{Khachatryan:2010xn}.

Jets and the missing transverse momentum in the event are determined from the four-vectors of
reconstructed particles, which are measured from the combination of the tracker and the
calorimeter informations~\cite{pflow}.
For jets, the anti-$k_{t}$ algorithm with a cone size of 0.5 is used \cite{antikt}. The transverse momentum of jets is required to be larger than 30 GeV with a pseudorapidity of $|\eta| < 2.5$.

An event is selected as a $\PW \rightarrow \ell \nu$ candidate if the following requirements
are satisfied:
(1) one isolated electron or muon
with a transverse momentum greater than 25 GeV and $|\eta| < 1.4$; events with
a second isolated electron or muon with a transverse momentum above 10~GeV are rejected;
(2) the missing transverse momentum (ascribed to the neutrino) greater than 30~GeV;
(3) the transverse mass = $\sqrt{ 2E_{T}(\nu) E_{T}(\ell) - 2 \vec{p}_{T}(\nu) \cdot \vec{p}_T(\ell)}$ of the charged lepton and the neutrino greater than 60 GeV.

Likewise, the following conditions are imposed to select $\cPZ(\gamma^{*}) \rightarrow \ell \ell $
(called $\cPZ$ in the following) candidates:
(1) two isolated electrons or muons
with opposite charge and each with a minimum transverse momentum of 25 GeV;
(2) at least one lepton with $ |\eta | < 1.4$;
(3) the reconstructed invariant mass of the dilepton system
between 60 and 120 GeV.

This selection resulted into essentially background-free $\PW$ and $\cPZ$ event samples.
The background for the $\PW$ sample is estimated to be less than 1\% and even smaller for the $\cPZ$ sample, independent of any additional requirements on the energies in the HF calorimeters.

The main features of the $\PW$ and $\cPZ$ event samples are found to be insensitive to
small variations of the selection criteria.
In the following, no efficiency corrections are applied,
and only direct comparisons to MC predictions are shown.

\subsection{Pileup rejection and single-vertex selection of \texorpdfstring{$\mathbf{W}$ and $\mathbf{Z}$}{W and Z} events}
  \label{sec:PUrejection}

As mentioned earlier, there can be several simultaneous $\Pp\Pp$ interactions in the same bunch
crossing in addition to the selected $\PW$ and $\cPZ$ events, the so-called pileup.
As the instantaneous luminosity increased steadily during the 2010 $\Pp\Pp$ data taking, the analysis was affected by very different pileup conditions.
For this analysis the data have been separated into three periods (P) with average instantaneous luminosities of
$ L_{inst} \leq 0.17 \; \mu$b$^{-1}$/s (P I),    $ 0.17 \; < L_{inst} \leq 0.34 \; \mu$b$^{-1}$/s (P II), and
$L_{inst} > 0.34 \; \mu$b$^{-1}$/s (P III), respectively.
Assuming a total inelastic cross section of about 70 mb~\cite{Aad:2011eu,Antchev:2011zz}, an instantaneous luminosity of $0.17 \; \mu$b$^{-1}$/s corresponds to an average of about one inelastic
pileup event.

The selection efficiency for $\PW$ and $\cPZ$ events is independent of the instantaneous luminosity. However, the charged-particle multiplicities and the energy depositions in the forward region of the detector (HF calorimeters), and thus especially the LRG signature, are strongly affected by the pileup (i.e., the gap is filled in).

The effects from pileup events for our analysis have been studied by means of zero-bias data samples
where the only requirement was that of colliding beams in the detector. Such event samples were
collected and analyzed for different instantaneous luminosities and the different periods. More details
are given in Section 5.1.

In order to limit the consequences of pileup, events with more than one vertex are rejected.
For this analysis a primary vertex, the $\PW(\cPZ)$-vertex, is defined as the one which contains
the lepton track(s). Events with additional vertices, formed by at least three tracks, are rejected.
Details about the vertex reconstruction algorithm are given in
Ref. \cite{CMS-Track-performance}.
The $\PW$-vertex $z$-position distribution is roughly Gaussian with a mean of 0.52~cm and a standard deviation of 5.9~cm.

The observed increase in the number of vertices has been studied using a zero-bias event sample
and found to be in agreement with the number of pileup events expected on the basis of the luminosity increase (cf.~Table 1).
pileup events can be categorized as hard and soft events. Hard $\Pp\Pp$ pileup interactions have some detectable charged particles in the central region of the detector
and are removed by the multiple-vertex veto.
The soft component has little or no detectable transverse activity in the central region and
does not result in reconstructed vertices.

The efficiency to reconstruct pileup vertices (including vertex splitting) has been
determined from MC simulations of minimum-bias and $\PW \to \Pe \nu$ events.
Based on the pileup conditions in the 2010 data, the efficiency to detect $\Pp\Pp$ pileup interactions
was found to be essentially constant within $\pm 25$ cm along the z direction of the nominal
interaction point with an average of about 72\%. The inefficiency essentially depends only on the
amount of soft pileup interactions (e.g., events without detectable charged particles in the central region of the detector)
in the MC simulation.
The reconstruction efficiency for pileup vertices was found to be essentially independent for the different luminosity periods, as long as the vertices were separated by more than 0.1 cm.
The corresponding inefficiency to detect the merged pileup vertex is estimated to be 3.3\%.
The uncertainties from the remaining soft pileup events on the results presented below are
discussed in more detail in Section 5.1.

The numbers of single-vertex $\PW$ and $\cPZ$ events are summarized in Table 1.
The single-vertex event yields, compared to the inclusive event yields, decrease
with increasing instantaneous luminosity and are in agreement with the expected numbers of vertices
from the simulation, assuming Poisson distributions.

\small
{
\begin{table*}[h]
\centering
\caption{Number of $\PW$ and $\cPZ$ candidate events with a single primary vertex. The numbers are given for the total and the three data-taking periods of different instantaneous luminosities. The percentage of
single-vertex events with respect to all selected $\PW$ and $\cPZ$ candidates for that period
is given in parentheses.
}
\begin{tabular}{|c|c|c|c|c|}
\hline
Single-vertex events      & $\PW \to \Pe \nu$    & $\PW \to \mu \nu$    & $\cPZ \to \Pe\Pe$   & $\cPZ \to \mu \mu$ \\
\hline
Total   &  13995  (25.8\%)  &   17924 (26.2\%)    &  1749  (25.7\%)   &  2924 (26.1\%) \\
P I      &   1502  (55.0\%)  &   1926 (53.0\%)  &  188 (53.0\%)  &   328 (56.8\%) \\
P II     &   7961  (31.0\%)          &    10524  (32.7\%)        &  1018 (31.4\%)    &  1718 (31.0\%) \\
P III    &   4532  (17.6\%)         &      5474   (17.3\%)      &    543  (16.8\%)   &    878 (17.2\%) \\

\hline
\end{tabular}\vspace{0.3cm}
\end{table*}
}
\normalsize

\section{Central charged-particle multiplicity and forward energy flow}
\label{sec:trackmult_energyflow}

The observables used to study the underlying event structure in $\PW$ and $\cPZ$ events are
the charged-particle multiplicity in the central detector, the energy depositions in both HF
calorimeters (in the following designated as HF$+$ and HF$-$ depending on the sign of the corresponding
$\eta$ coverage),
and the correlations between them. In the following, only the distributions for $\PW \to \ell \nu$
events are discussed. No significant differences are observed between $\PW$ events selected with decays to electrons or muons.
The same analysis has been performed on the
$\Pp\Pp \rightarrow \cPZ X$ data sample and consistent results are obtained.

For the studies
described below, the charged-particle multiplicity is measured in the range $|\eta| < 2.5$ for
track momentum thresholds of $p_{T} > 0.5$ GeV and $p_{T} > 1.0$ GeV, excluding
the tracks associated with $\PW$ decays. Additionally, in order to study the underlying event structure,
events with central jet activity ($p_{T} > 30$ GeV,  $|\eta| < 2.5$) are excluded from the multiplicity plots.

A detailed study of the pion track reconstruction within
the acceptance of the tracker \cite{CMS-Track-performance} determined
that the efficiency rises from about 88\% at a $p_{T}$ of  0.5 GeV to about 95\%
for $p_{T}$ between 1--10 GeV. Above 10 GeV, the efficiency decreases slowly to about 90\%
at 50 GeV. Furthermore, it was shown that the hadron track reconstruction efficiency in the data
agrees within 1-2\% with the one in the MC simulation. The total
systematic uncertainty of the tracking efficiency was estimated to be less than 3.9\%.
The observed charged-particle multiplicities, excluding the lepton(s) from the $\PW(\cPZ)$ decays,
vary between 0 and about 50,
with an average of 11 and an r.m.s.~of 8.2 for $p_{T} > 1.0$ GeV.
About twice as many tracks are found with the lower threshold, $p_{T} > 0.5$ GeV, and
about 0.15\% of the events have more than 100 tracks.
Nearly identical charged-particle multiplicity distributions are observed in electron- and muon-tagged
events with the same $p_{T}$ thresholds. The charged-particle multiplicity distribution
for $\PW \to \Pe \nu $ events is shown in Fig. 2a for tracks with
$p_{T} > 1.0$~GeV, and for $\PW \to \mu \nu $ events with for $p_{T} > 0.5$~GeV in Fig. 2b.
The corresponding distributions for $\PW \to \Pe \nu $ with $p_{T} > 0.5$~GeV and $\PW \to \mu \nu $
with $p_{T} > 1.0$~GeV are consistent with the displayed distributions in Figures 2a and 2b.

\begin{figure*}[hbtp]
\subfigure[]{\label{fig:eAllTracks}\includegraphics[width=0.5\linewidth]{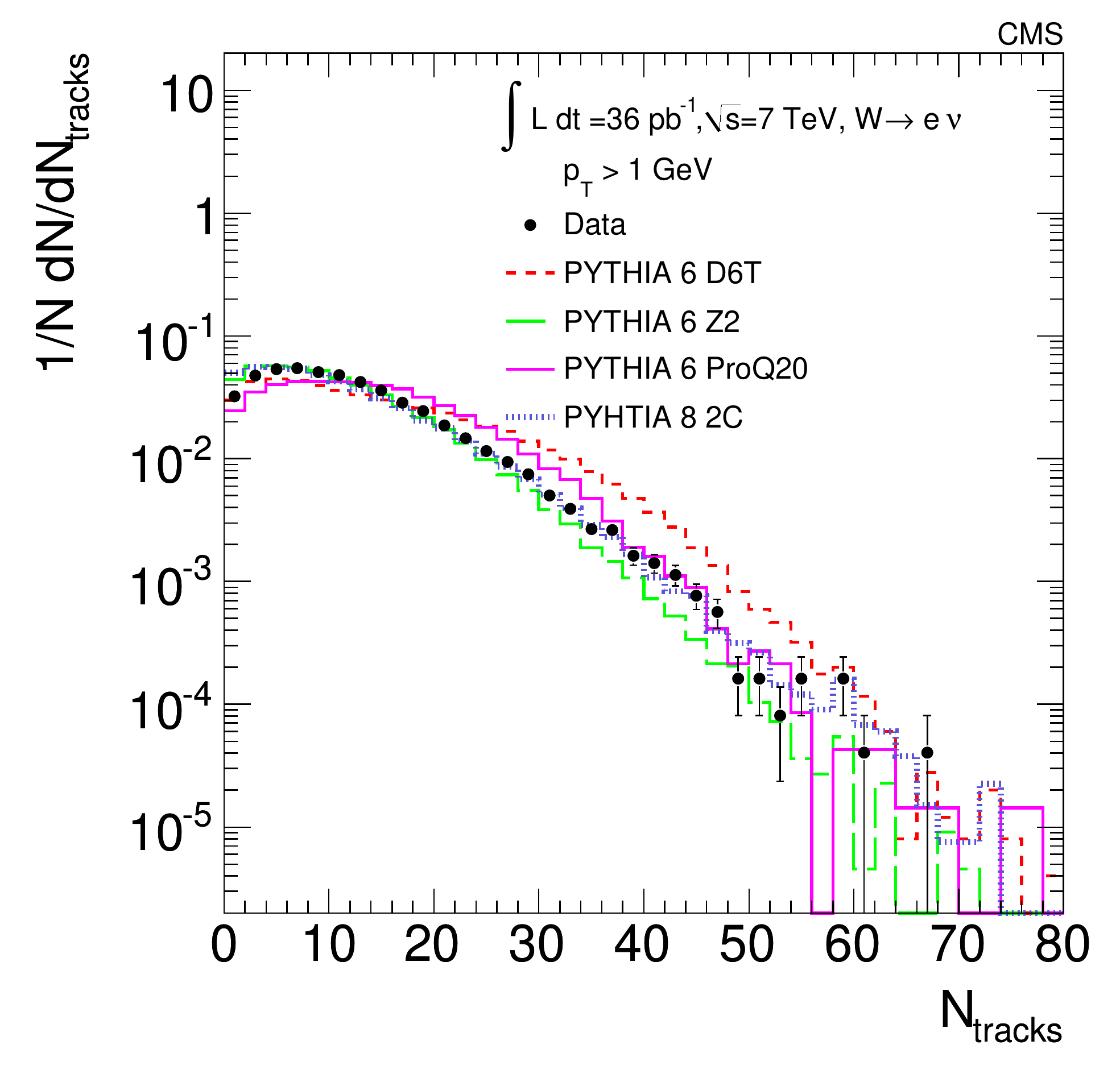}}
\subfigure[]{\label{fig:muAllTracks}\includegraphics[width=0.5\linewidth]{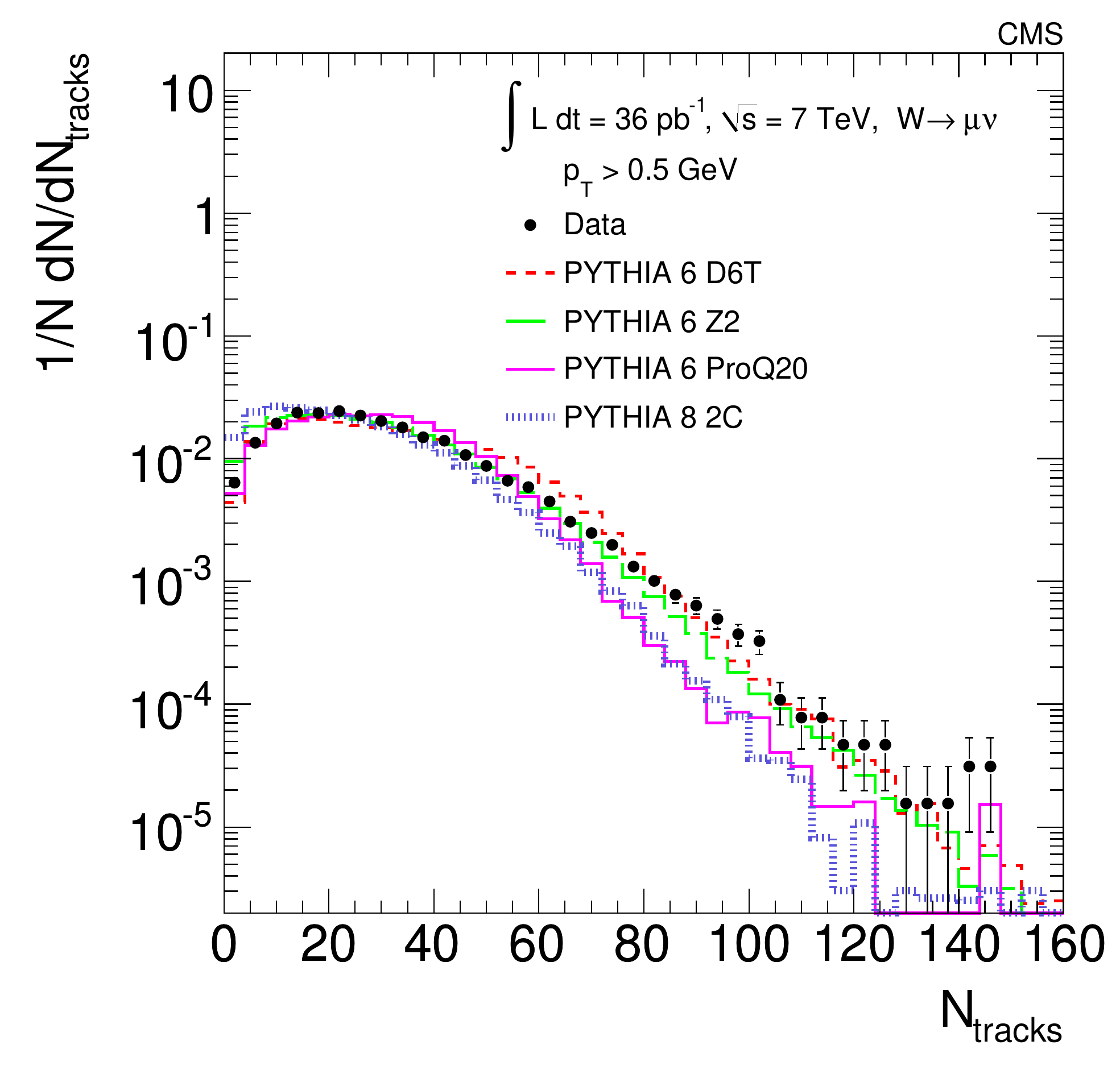}}  \\
\subfigure[]{\label{fig:eSumEHF}\includegraphics[width=0.5\linewidth]{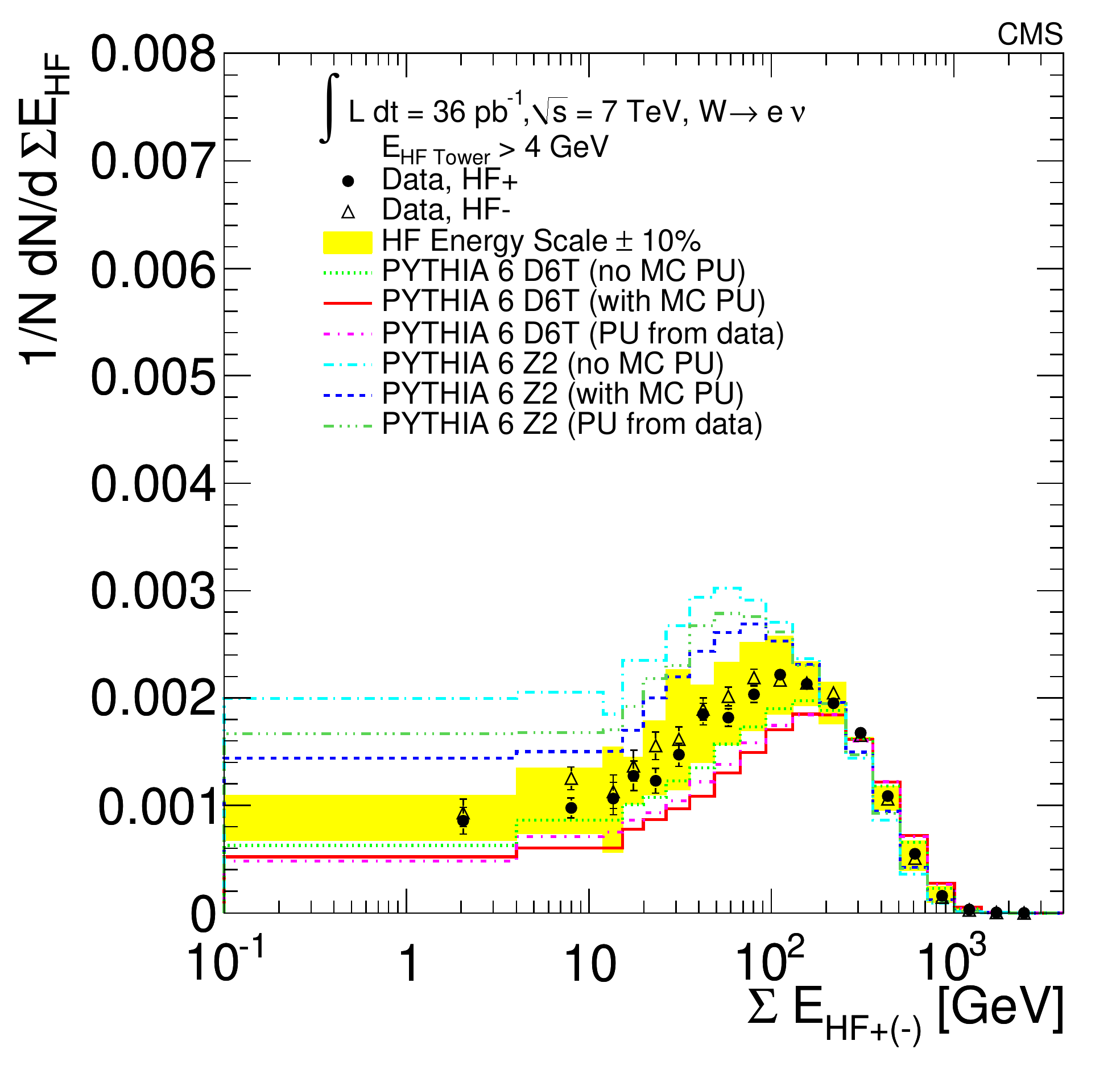}}
\subfigure[]{\label{fig:muSumEHF}\includegraphics[width=0.5\linewidth]{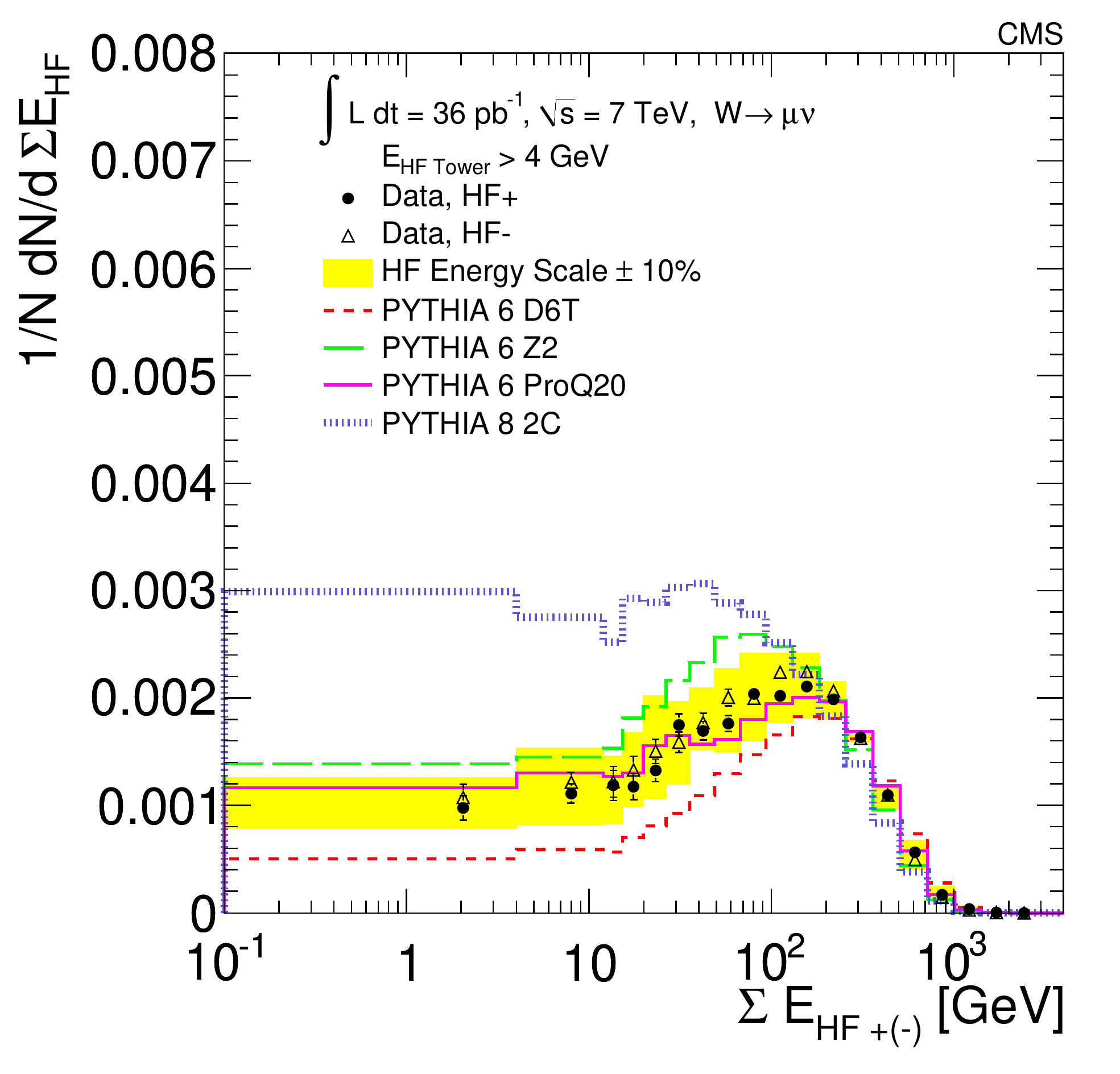} }

\caption{
The charged-particle multiplicities and the summed HF$+$ and HF$-$ energy distributions
in a) and c) $\PW \rightarrow \Pe \nu X$
and b) and d) $\PW \rightarrow \mu \nu X$
candidate events
are shown for data and MC simulations, including pileup, with different tunes for the underlying event.
The uncorrected charged-particle multiplicities are shown for electron- and muon-tagged $\PW$ events, for two thresholds on
track transverse momenta of (a) $p_{T} > 1.0$~GeV
and (b) $p_{T} > 0.5$~GeV
The band shown for the HF energy distributions indicates the uncertainty related to a $\pm$ 10\% HF energy scale variation. The effects of the pileup on the MC simulation can be seen in (c) where the
MC HF energy distributions are shown with pileup taken either from data or from the MC simulation
and without pileup.
}
\label{Fig2-mult}
\end{figure*}

The \PYTHIAeight generator with tune 2C provides the best overall description for the higher track $p_{T}$
threshold, and \PYTHIAsix with tune Z2 provides a reasonable description for both track $p_{T}$
thresholds. However, both \PYTHIAeight 2C and \PYTHIAsix Z2 predict too many events
with very small charged-particle multiplicities.
The \PYTHIAsix D6T tune predicts a harder-$p_{T}$ spectrum
for hadrons in the underlying event and thus a larger multiplicity in the case of the higher threshold.
The Pro-Q20 tune of \PYTHIAsix significantly
underestimates the event yields with very high multiplicities (Fig. \ref{fig:muAllTracks}).

The energy deposition in the HF$+$ and HF$-$ calorimeters
is determined from the sum of individual calorimeter towers with an energy
threshold of 4~GeV, corresponding to a minimum transverse momentum of 0.07--0.4~GeV.
The uncertainty on the energy scale of the HF calorimeter was estimated to be
about $\pm10$\% (for details see \cite{pflow}). This uncertainty was taken into account by a $\pm10$\% scaling of the single-tower energy, resulting in new estimates of the total energy deposition in HF for the data, while keeping the MC unchanged.
The corresponding systematic uncertainty for the energy measurement in the HF is shown as a band in Fig. \ref{Fig2-mult} and all the following figures.
This uncertainty is much larger than the 3.5\% difference
between the reconstructed energy distributions in the HF$+$ and HF$-$ calorimeters (cf.\ Table \ref{Table2}).

The observed HF energies vary between 0~GeV (i.e., no HF tower with
an energy above 4~GeV) and more than 2~TeV.
Only a few events have an energy deposition above 2~TeV (about 0.05\% for electrons and muons
combined) and the highest energy deposition is 2.7~TeV. All events with high-energy
depositions also have large tower multiplicities.
The average energies observed in HF$+$ and HF$-$ are measured in 11 HF rings of calorimeter towers,
each covering approximately a $\Delta |\eta|$ range of 0.175.
The difference between the average energy deposition per ring ($\eta$ bin)
in the data and the different MC tunes is consistent for all tower rings.
The average energy deposited per $|\eta|$ ring predicted by the D6T (Z2) tune is too large (too small),
while the Pro-Q20 tune provides a very good description of the data.

The distributions of the total HF$+$ and HF$-$ energy sums ($E_{HF+(-)}$) are shown in Fig.\ \ref{fig:eSumEHF} for the data and the \PYTHIAsix D6T and Z2 tunes, with and without pileup (PU) in $\PW \rightarrow \Pe \nu$ events.
In order to take into account effects not included in the simulation (e.g., beam-gas interactions),
the soft pileup contribution obtained from the zero-bias data is added to the MC simulation without pileup.
It can be seen that the pileup contribution,  as determined from the data,
agrees very well with the one obtained from the luminosity-dependent MC simulation. Therefore, in the following analysis and all shown distributions,
the pileup contribution is estimated from the MC simulation.
Figure \ref{fig:muSumEHF} shows the HF$+$ and HF$-$ energy sums
for $\PW \rightarrow \mu \nu$ events and for several different MC tunes of the underlying event.

The mean values of the reconstructed HF energy and the tower multiplicities are
given in Table \ref{Table2}, again for the three different instantaneous luminosity periods and separately
for $\PW \rightarrow \Pe \nu$ and  $\PW \rightarrow \mu \nu$ events.

{
\begin{table*}[h]
\centering
\caption{Mean energy depositions and tower multiplicities in each HF for
single-vertex $\PW$ events from the three running periods with different instantaneous luminosities.
}
\begin{tabular}{|c|c|c|c|c|}

\hline

\hline

Mean energy [GeV]  & $\Pp\Pp \rightarrow \PW(\rightarrow \Pe \nu) X  $  & $\Pp\Pp \rightarrow \PW(\rightarrow \mu \nu) X  $ \\
                  \hline
HF$+$ / HF$-$, P I    &  298.3  / 284.7            & 295.7 / 286.4              \\
HF$+$ / HF$-$, P II    & 308.2 / 296.6             & 313.0 / 295.7             \\
HF$+$ / HF$-$, P III   & 322.6 / 313.6            &  329.6 / 310.8            \\
\hline
Average tower multiplicity & & \\ \hline

HF$+$ / HF$-$, P I            &  29.2  / 28.5          & 29.4 / 28.5  \\
HF$+$ / HF$-$, P II           &  30.5 / 29.6           & 30.9  / 29.6  \\
HF$+$ / HF$-$, P III          &  32.1 / 31.2           & 32.7 /  31.2  \\

\hline
\end{tabular}\vspace{0.3cm}

\label{Table2}
\end{table*}

The average HF energy deposition, averaging the energy deposits in the HF+ and HF- calorimeters,
in the data is 310 GeV, with an r.m.s. of 235 GeV.
On average, about 30 towers with more than 4 GeV are reconstructed in each HF calorimeter.
The statistical uncertainties of the mean energy values (mean tower multiplicities), estimated from the r.m.s. of the distribution, amount to  less than $\pm 5$ GeV ($\pm 0.4$) in the data and even smaller in the MC
simulation.
The corresponding mean value obtained with \PYTHIAsix D6T, including the HF energy
depositions from simulated pileup, is 370 GeV, with a tower multiplicity of 35.
The \PYTHIAsix Z2 tune predicts a mean energy deposition of 270 GeV
and a tower multiplicity of 27, whereas using the Pro-Q20 tune results in a simulated
energy deposition of 311 GeV and a tower multiplicity of 29 towers, similar to the data.

As can be seen from Figs. \ref{fig:eSumEHF} and  \ref{fig:muSumEHF}, besides the Pro-Q20 tune,
none of the MC models considered provide a good description of the HF energy distribution
observed in the data. For energy depositions between 10 and 150 GeV, large differences
between the data and different tunes
are observed. In particular, the number of events in the data is about 30 to 50\% higher than predicted by the
D6T tune, and 50\% lower than predicted by the Z2 tune.
For simplicity, the older P0 and Pro-PT0 tunes are omitted from the following more detailed studies.

In total, 287 $\PW$ and $\cPZ$ events with no individual tower energy deposition above 4 GeV
in one HF calorimeter, are observed.
These events are defined as LRG events, i.e., events with ``zero" energy depositions, and
are discussed in detail in Section 6.

\subsection{Soft pileup events and HF energy distributions}
\label{sec:soft_pile_up}

The observed mean energy values in the HF increased by about 10 $\pm$ 5 GeV from period I to period II and by about 15 $\pm$ 5 GeV from period II to period III (cf.\ Table \ref{Table2}), both in
$\PW \rightarrow \Pe \nu$ and  $\PW \rightarrow \mu \nu$ events. This increase of HF energy depositions is interpreted as arising from soft pileup events, not identified by the vertex finder.
A similar increase of the mean energy deposition in HF is also seen in the MC simulations, when events
with and without pileup are compared.

The properties of such soft pileup events have been studied with the zero-bias data samples,
where the only requirement was that of colliding beams in the detector, taken during the different running periods.
In events from this sample with zero reconstructed vertices, three classes of events can be identified:
(1) events with no energy deposition in either HF (hereafter referred to as quasi-elastic $\Pp\Pp$-scattering),
(2) events with zero energy in only one of the HF calorimeters
and non-zero energy in the other (soft scattering with a LRG signature), and
(3) events with non-zero energy depositions in both HF calorimeters (soft inelastic $\Pp\Pp$-scattering).

The contributions of beam-gas events and other beam-related backgrounds to the
HF energy dispositions were studied in randomly triggered events with non-colliding beams
and were found to be negligible.

The relative fraction of quasi-elastic $\Pp\Pp$-scattering event candidates in the zero-bias samples
decreases from about 50\% in period I to 20\% in
period III, while the fraction of soft inelastic events increases from about 15\% in period I
to 40\% in period III.
The fraction of soft events with a LRG signature of about 40\% is roughly constant across the three luminosity periods.

In conclusion, soft pileup events, not identified by the single-vertex requirement, can have an important
effect on the HF energy distributions and on LRG events in particular. As discussed below, their
contribution is well modeled by the MC simulations.

\subsection{Correlations of the forward energy flow and the central charged-particle multiplicity}

In the following, the correlation between the central charged-particle multiplicity and the forward energy flow is measured in the data and compared to MC models.
For this study, events with energy depositions in the HF$-$ calorimeter of
20-100 GeV (low), 200-400 GeV (medium), and above 500~GeV (high) are selected.
The central charged-particle multiplicity distributions with track $p_{T}$ thresholds of  $1$ GeV (for $\PW\rightarrow \Pe\nu$ events)
and $0.5$ GeV (for $\PW\rightarrow \mu\nu$ events) are shown in Fig. \ref{Fig3-mult}, while the HF$+$ energy distributions
for the three HF$-$ energy intervals are shown in Fig. \ref{Fig4-HFdist} (for $\PW\rightarrow \Pe\nu$ and $\PW\rightarrow \mu\nu$ events).

The charged-particle multiplicity distributions for the medium HF$-$ energy range (Figs. \ref{fig:eMedTracks} and \ref{fig:muMedTracks}) are described reasonably well by the
\PYTHIAsix D6T and Z2 tunes. The agreement between the data and the D6T tune is poorer when a
$1$ GeV track $p_{T}$ threshold is applied.
For the low HF energy range (Figs. \ref{fig:eLowTracks} and \ref{fig:muLowTracks}), the D6T tune fails to describe the charged-particle multiplicity distribution, whereas the Z2 tune is in good agreement with the data, after applying a $0.5 \; \mathrm{GeV}$ track $p_{T}$ threshold.
Finally, when requiring a large HF energy deposition (Figs. \ref{fig:eHighTracks} and \ref{fig:muHighTracks}),
the Z2 tune provides a good description, whereas the D6T tune overestimates the charged-particle multiplicity.

\begin{figure*}[htp]
\centering
\subfigure[]{\label{fig:eLowTracks}\includegraphics[width=0.4\linewidth]{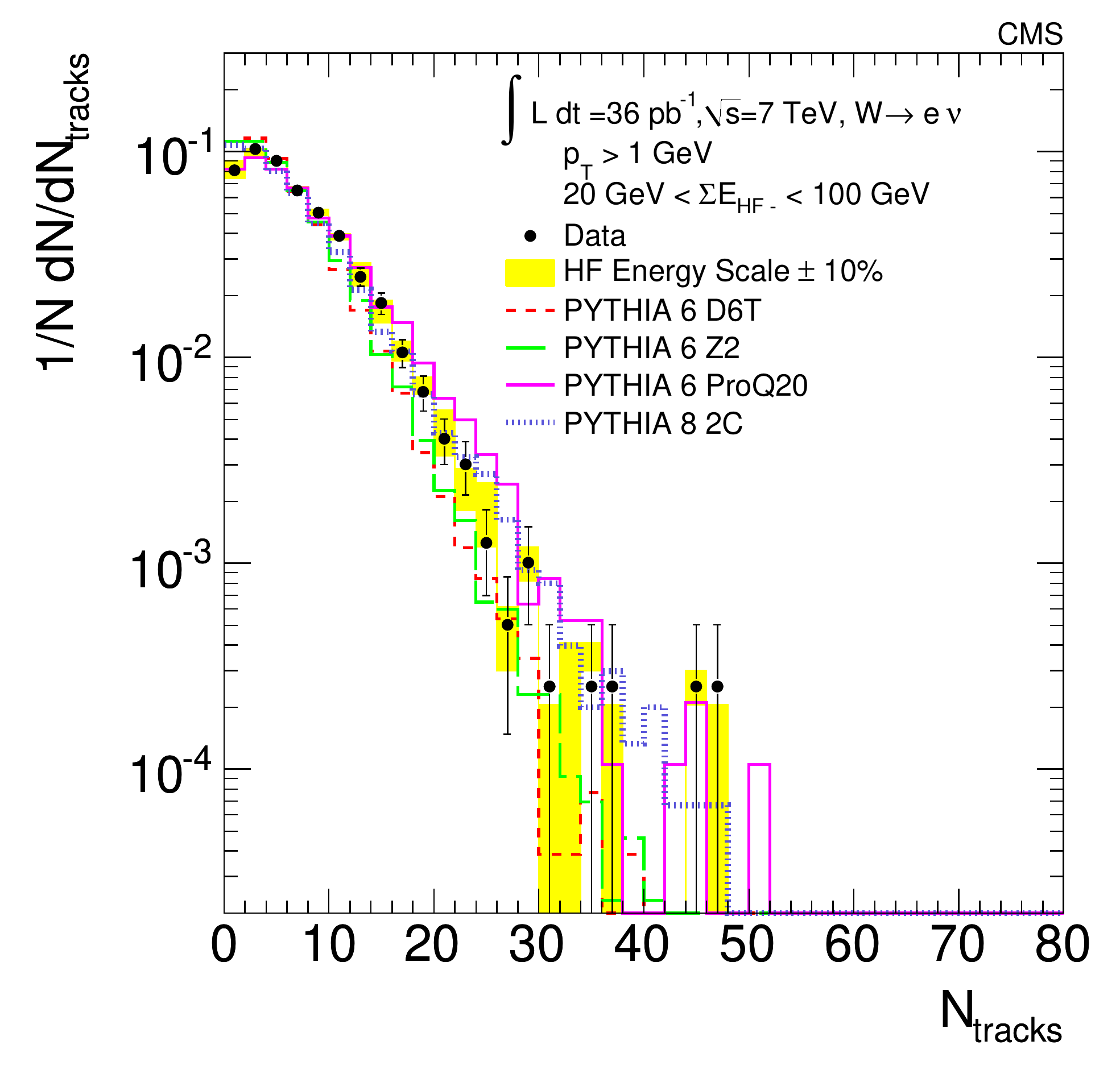}}
\subfigure[]{\label{fig:muLowTracks}\includegraphics[width=0.4\linewidth]{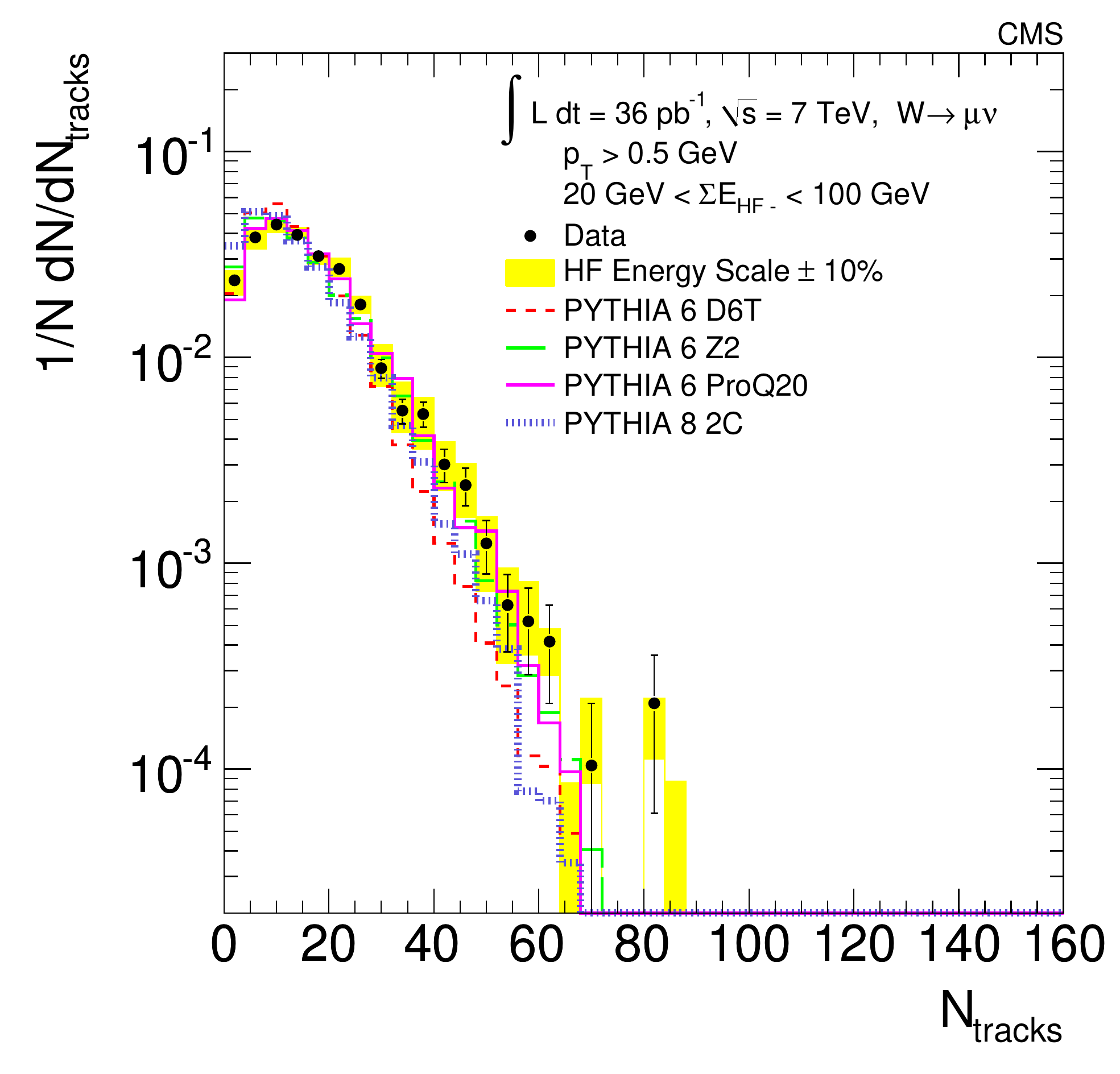}}

\subfigure[]{\label{fig:eMedTracks}\includegraphics[width=0.4\linewidth]{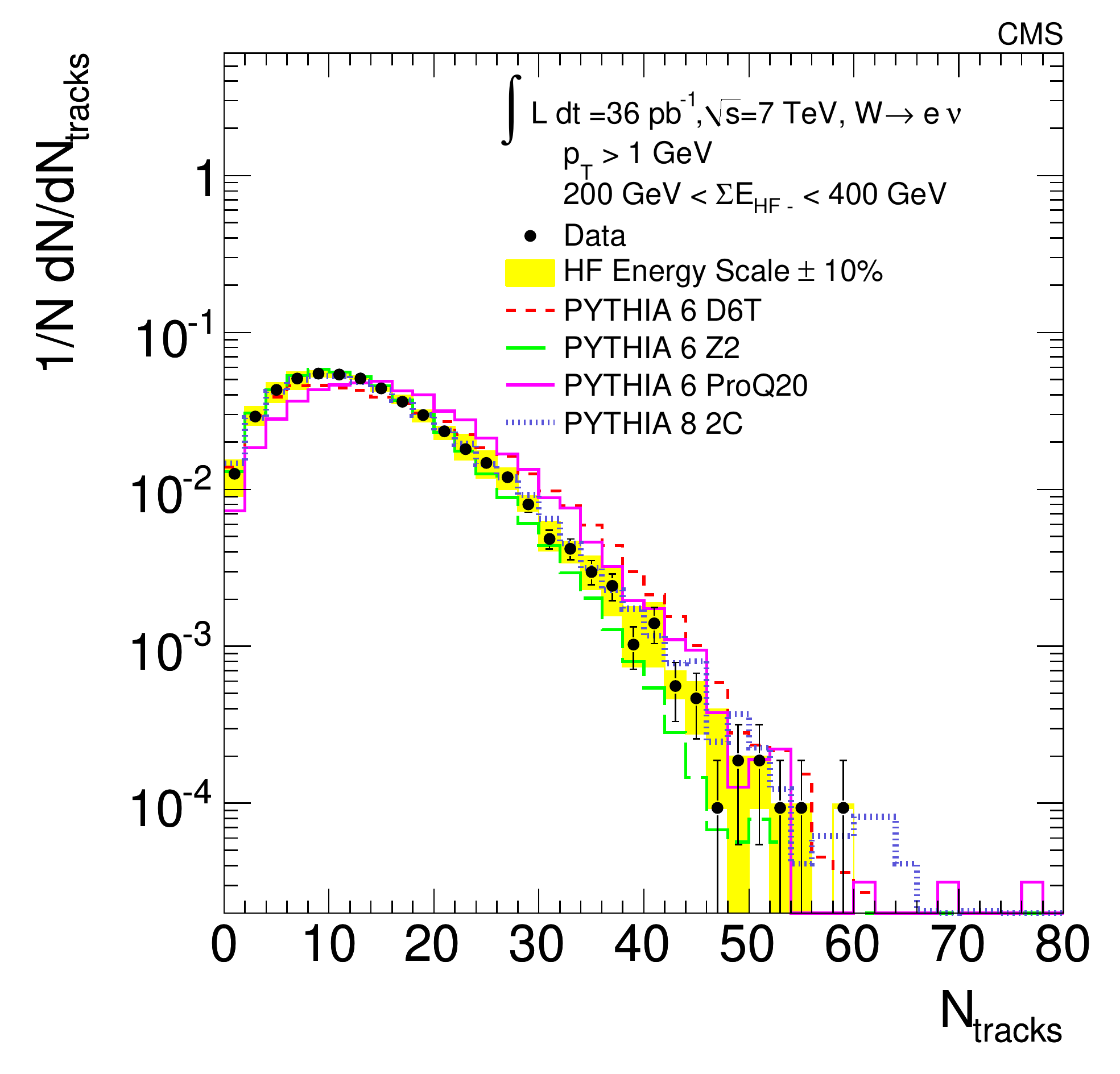}}
\subfigure[]{\label{fig:muMedTracks}\includegraphics[width=0.4\linewidth]{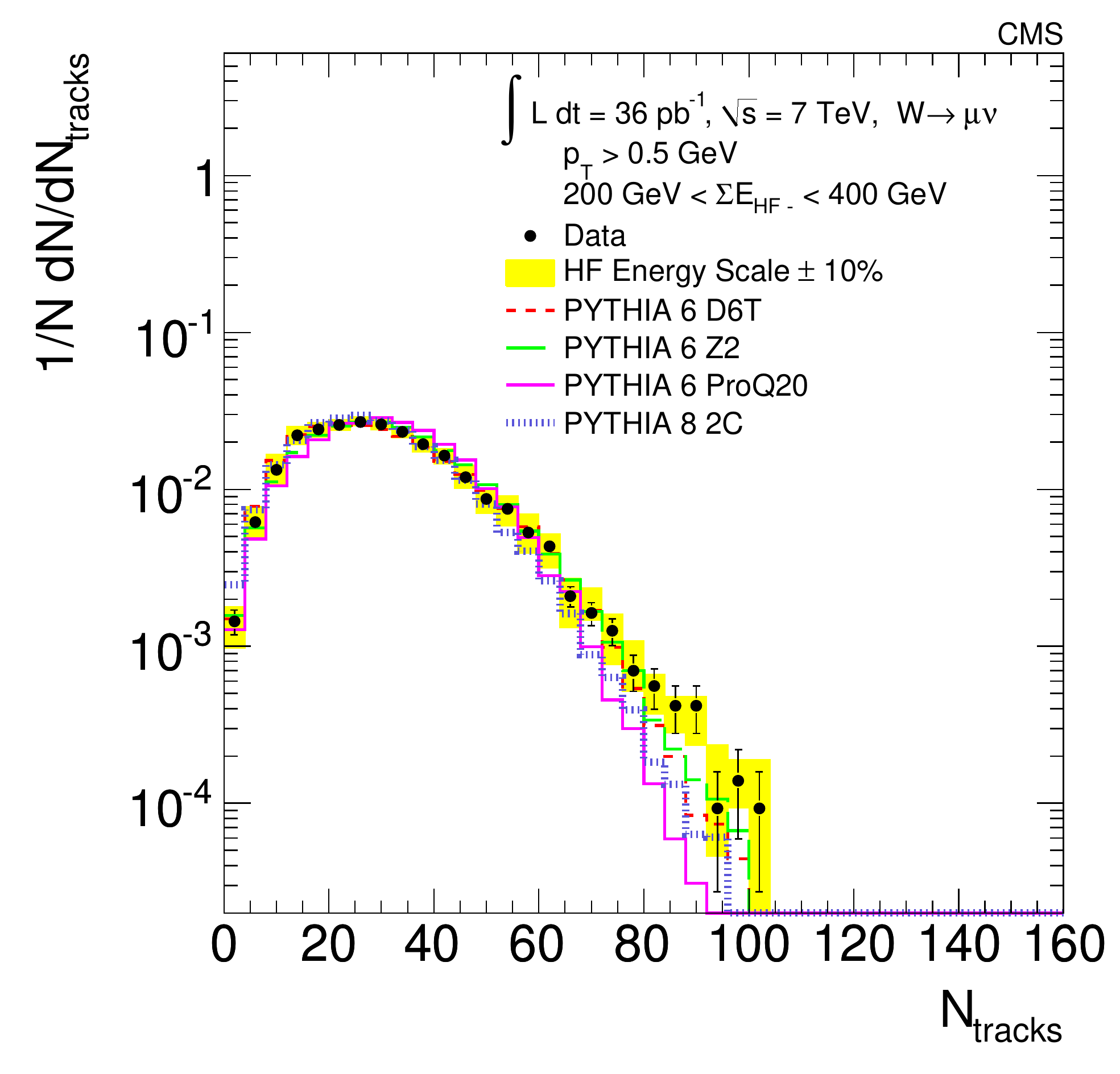}}

\subfigure[]{\label{fig:eHighTracks}\includegraphics[width=0.4\linewidth]{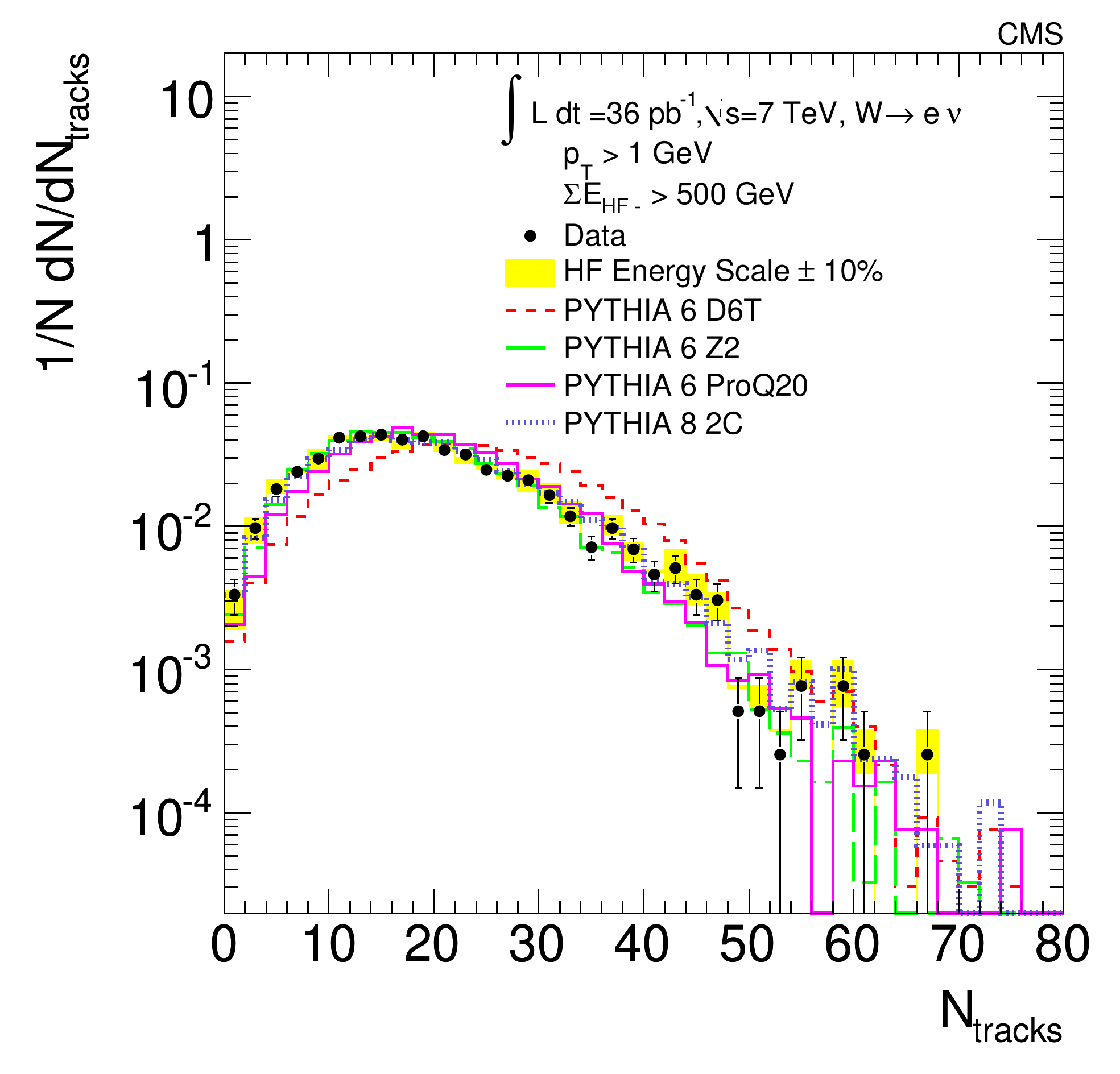}}
\subfigure[]{\label{fig:muHighTracks}\includegraphics[width=0.4\linewidth]{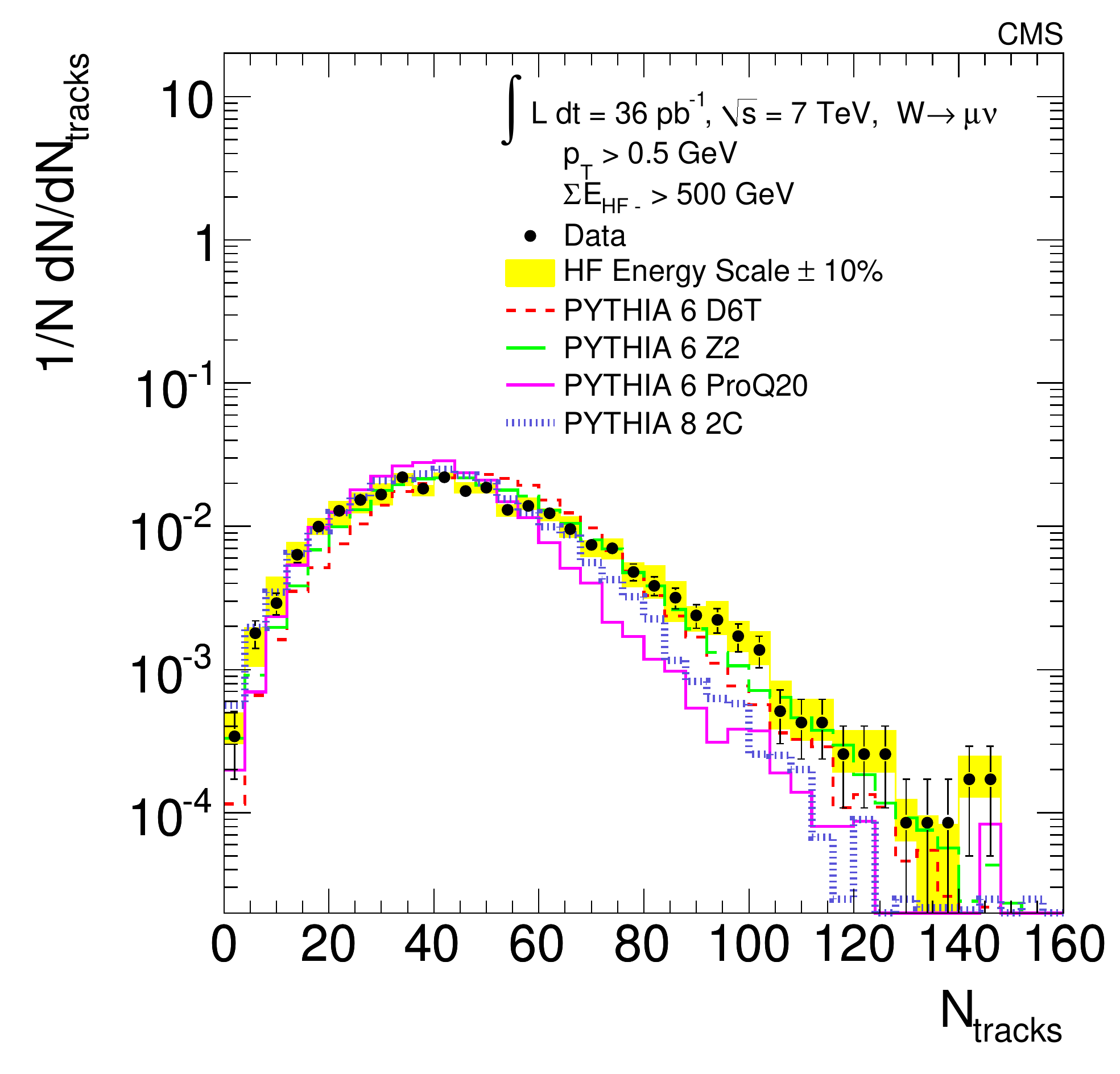}}

\caption{Charged-particle multiplicity distributions in the data and from MC simulations with different
tunes, for the three HF$-$ energy intervals
of (a) and (b)
20-100~GeV,
(c) and (d) 200-400~GeV,
and (e) and (f) $>500$~GeV.
The plots in the left column are for $pp \rightarrow W^{\pm}X \rightarrow e^{\pm} \nu X$, for tracks with $p_{T} > 1.0$~GeV,
and those in the right column for $pp \rightarrow W^{\pm}X \rightarrow \mu^{\pm} \nu X$,
for $p_{T} > 0.5$~GeV (different scales are used for the $x$-axes).
}
\label{Fig3-mult}
\end{figure*}

The HF$+$ distributions for the medium HF$-$ energy interval (Figs. \ref{fig:eMedSumEHF} and
\ref{fig:muMedSumEHF}) are in better agreement with the predictions of the various tunes than the inclusive HF distributions (Figs. \ref{fig:eSumEHF} and \ref{fig:muSumEHF}).
On the other hand, when requiring a low HF$-$ energy deposition, the HF$+$
energy distribution is poorly modeled by all MC tunes (Figs. \ref{fig:eLowSumEHF} and \ref{fig:muLowSumEHF}).
Finally,  for events with high-energy depositions in HF$-$, the \PYTHIAeight generator,
which in the inclusive
case underestimates the rate of events with large HF$-$ energy depositions,
provides a good description of the energy distribution in HF$+$.
Conversely, all other tunes predict more events with large HF$+$ energy than observed in the data (Figs. \ref{fig:eHighSumEHF} and \ref{fig:muHighSumEHF}).

\begin{figure*}[htp]
\centering
\subfigure[]{\label{fig:eLowSumEHF}\includegraphics[width=0.4\linewidth]{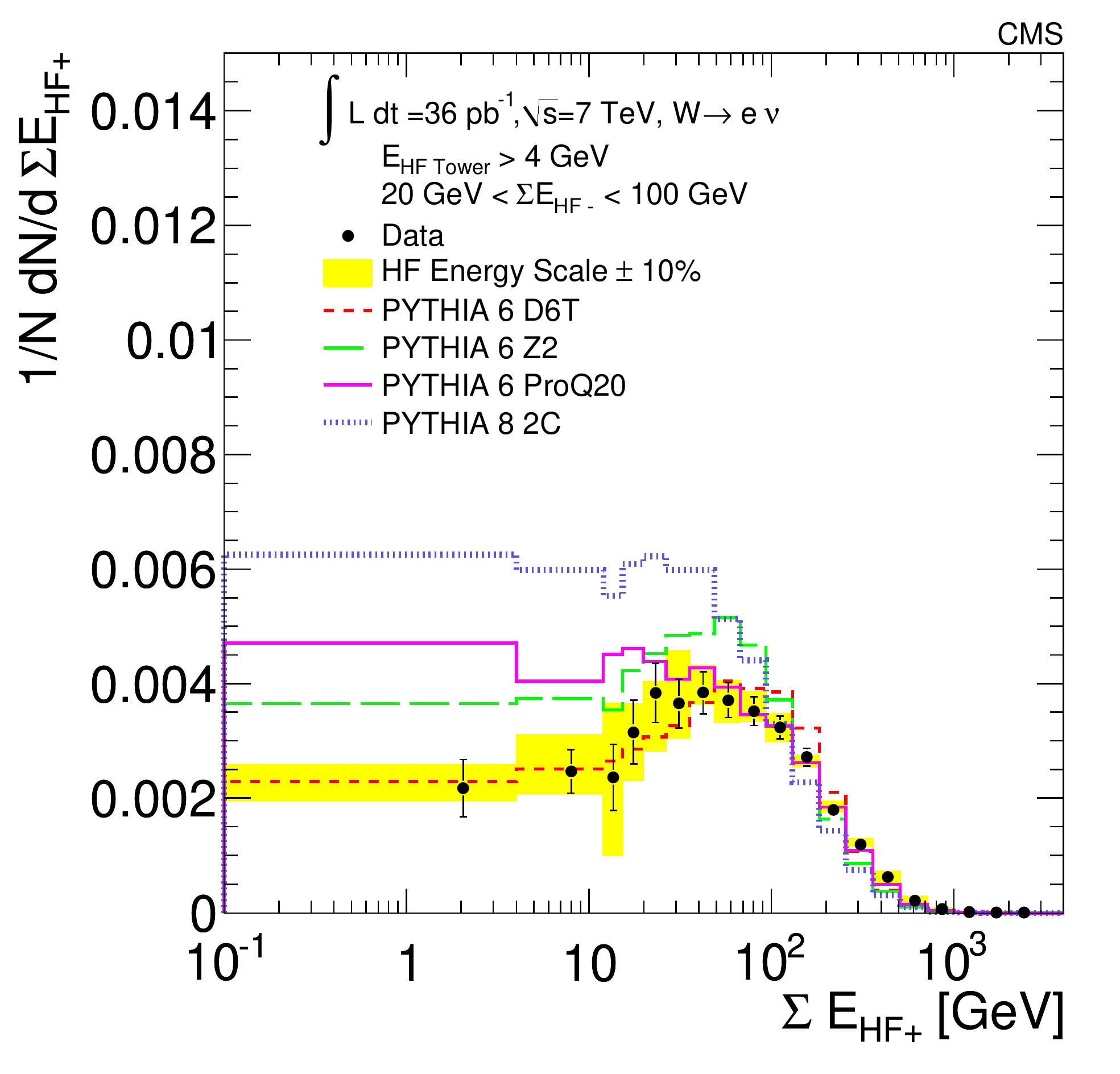}}
\subfigure[]{\label{fig:muLowSumEHF}\includegraphics[width=0.4\linewidth]{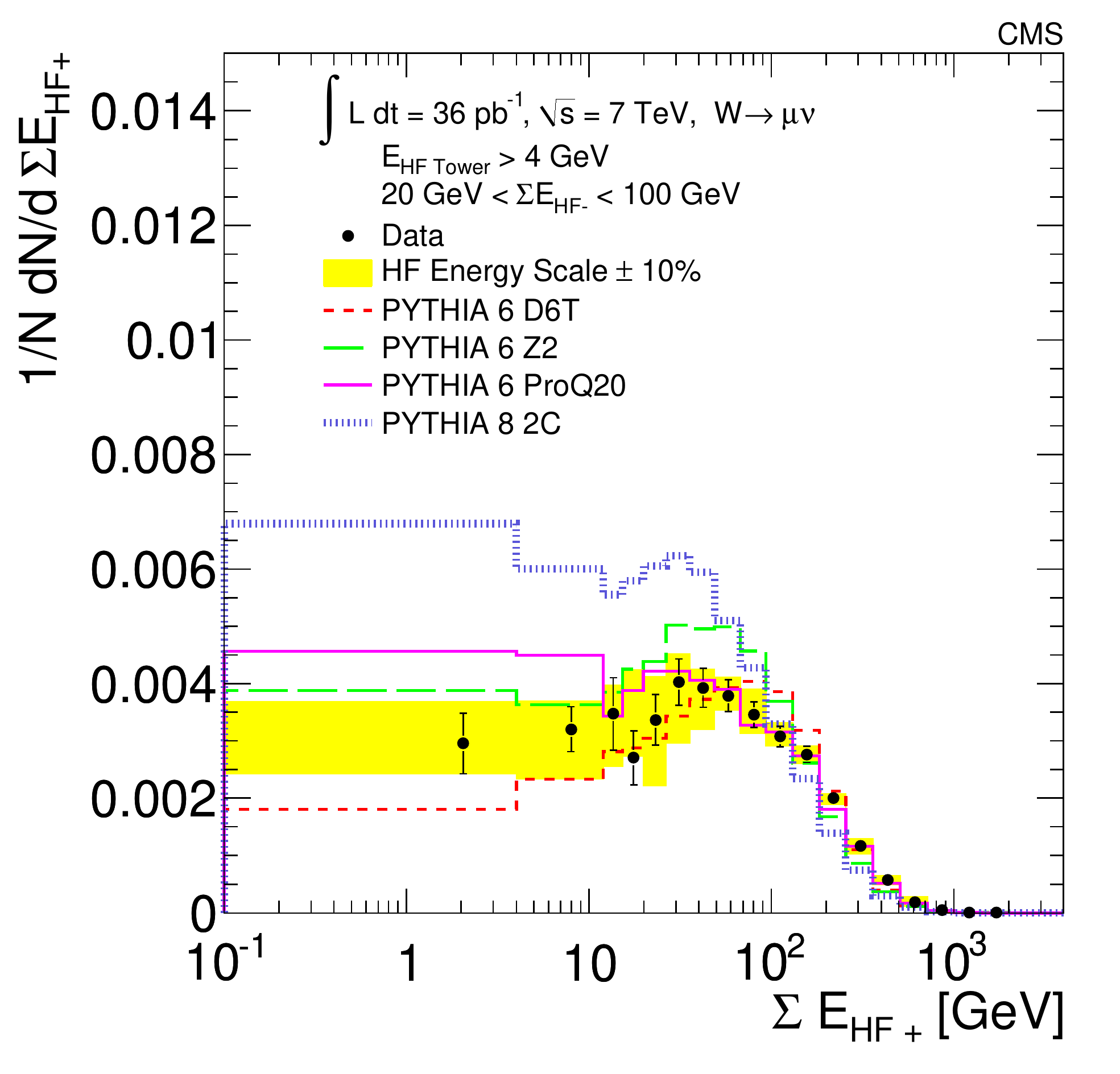}}

\subfigure[]{\label{fig:eMedSumEHF}\includegraphics[width=0.4\linewidth]{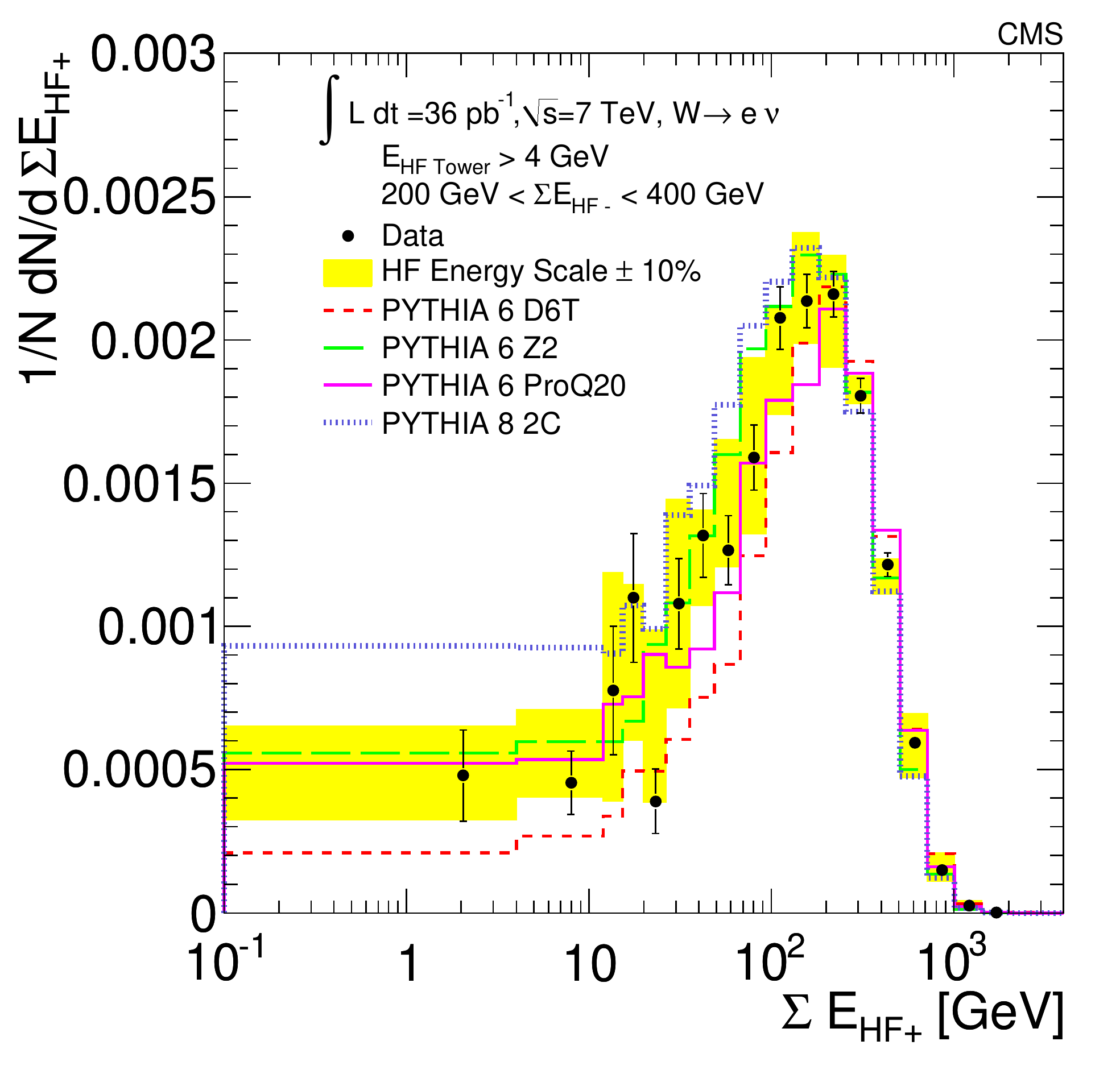}}
\subfigure[]{\label{fig:muMedSumEHF}\includegraphics[width=0.4\linewidth]{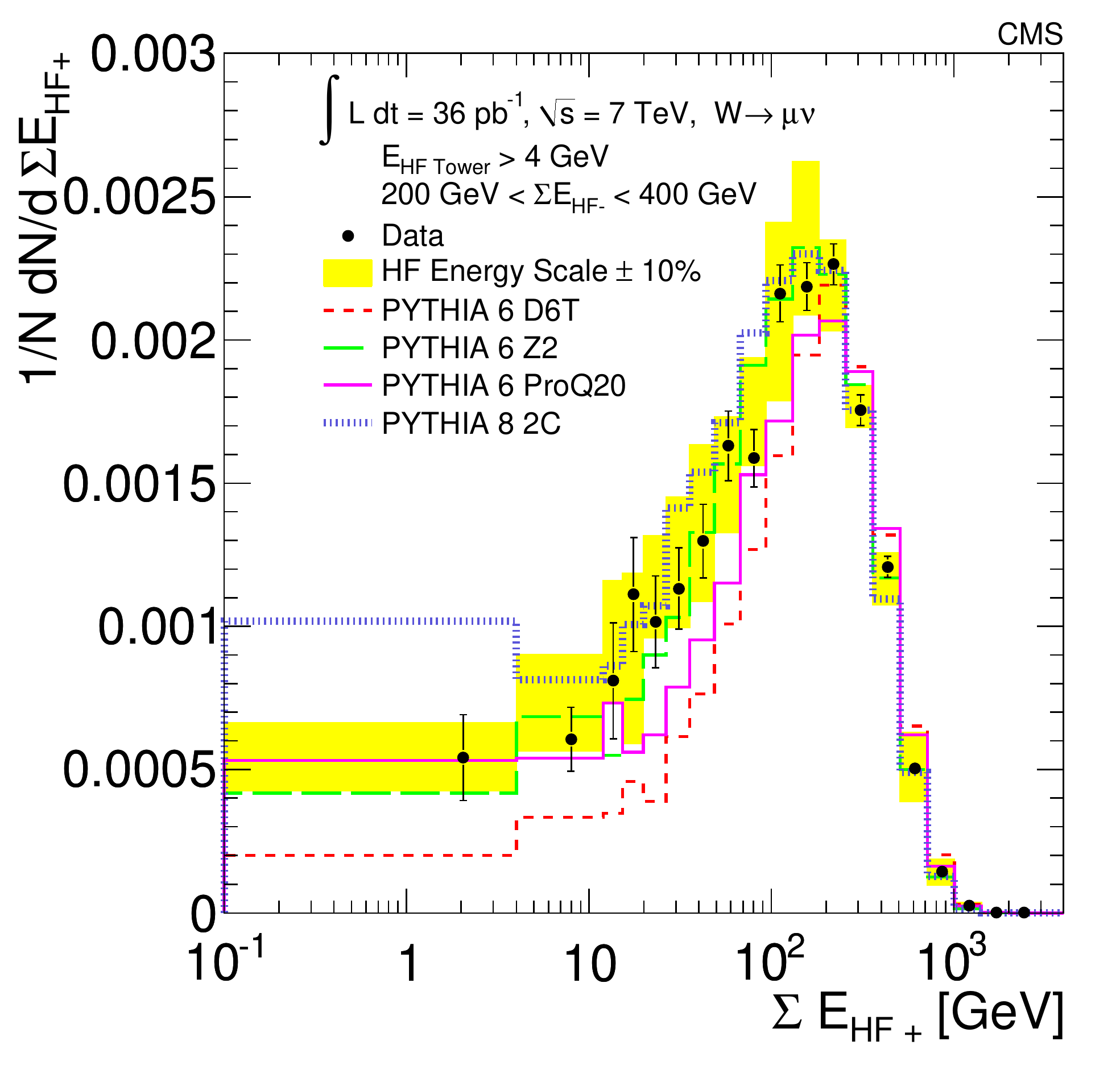}}

\subfigure[]{\label{fig:eHighSumEHF}\includegraphics[width=0.4\linewidth]{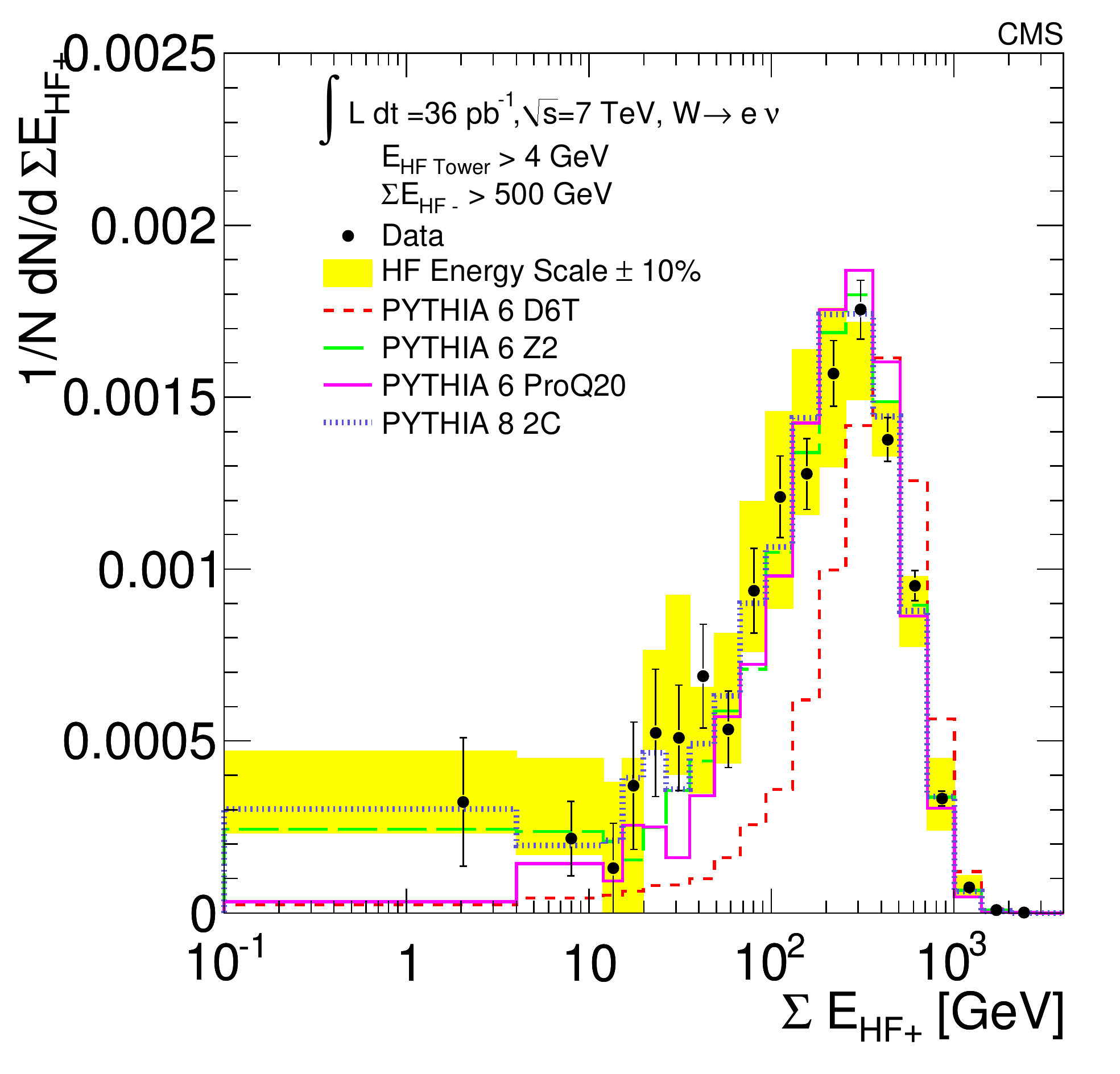}}
\subfigure[]{\label{fig:muHighSumEHF}\includegraphics[width=0.4\linewidth]{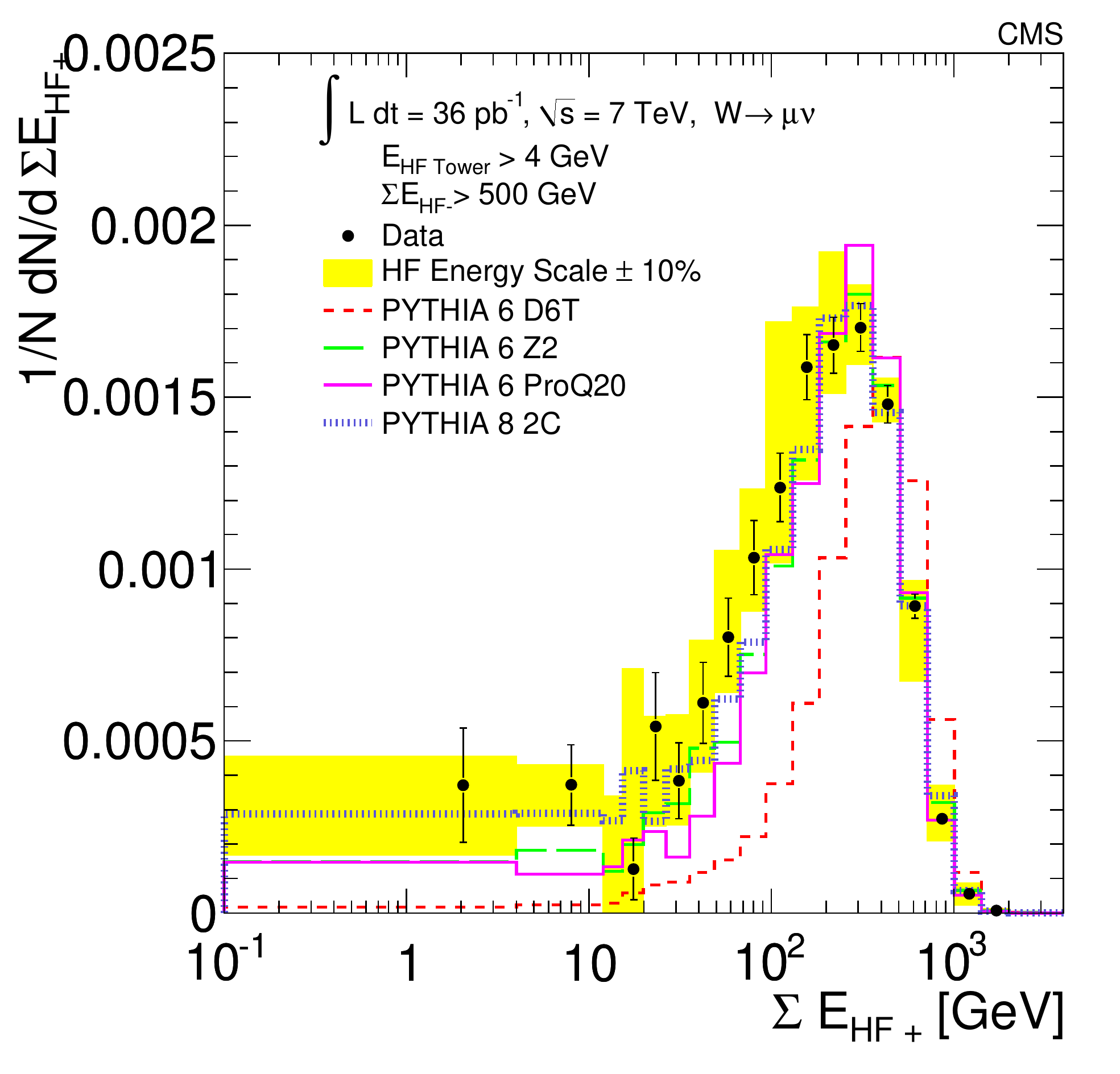}}

\caption{Energy distribution in the HF$+$ calorimeter, shown
for different HF$-$ energy intervals (a) and (b)
20-100 GeV,
(c) and (d) 200-400 GeV,
(e) and (f) $>500$ GeV),
for data and different MC tunes.
In the left column, the plots are shown for $\Pp\Pp \rightarrow \PW X \rightarrow \Pe \nu X$ events and in the right column, for $\Pp\Pp \rightarrow \PW X \rightarrow \mu \nu X$.
}
\label{Fig4-HFdist}
\end{figure*}

Figure \ref{Fig5-lowerhigher} shows the minimum and maximum energy depositions per event in the HF$+$ and HF$-$ calorimeters for the data and the various MC tunes.
In comparison to Fig. \ref{Fig2-mult}, the differences between the data and all available MC tunes are somewhat enhanced.

\begin{figure*}[hbtp]
\subfigure[]{\label{fig:eMinEHF}\includegraphics[width=0.5\linewidth]{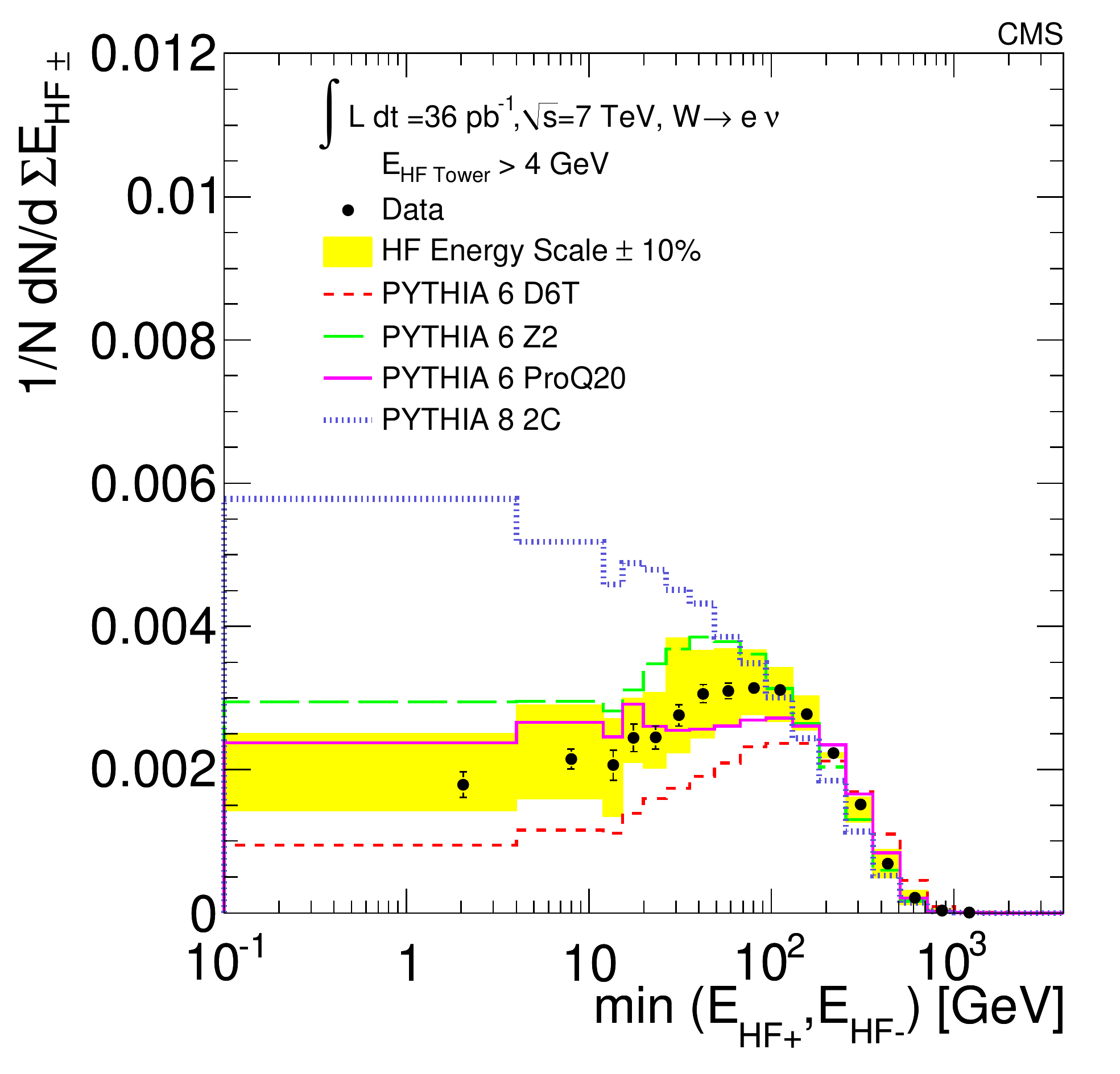} }
\subfigure[]{\label{fig:muMinEHF}\includegraphics[width=0.5\linewidth]{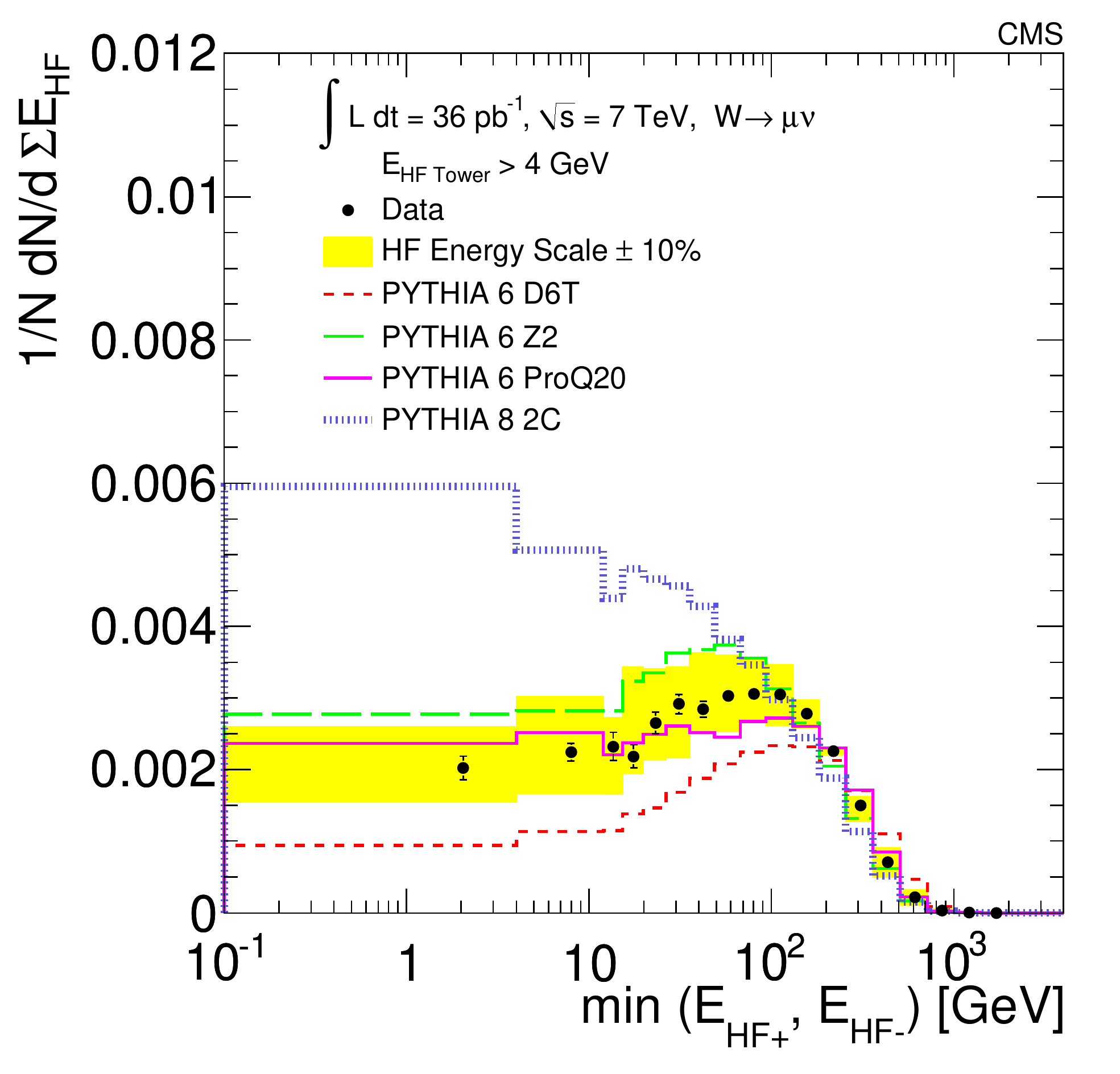}}  \\
\subfigure[]{\label{fig:eMaxEHF}\includegraphics[width=0.5\linewidth]{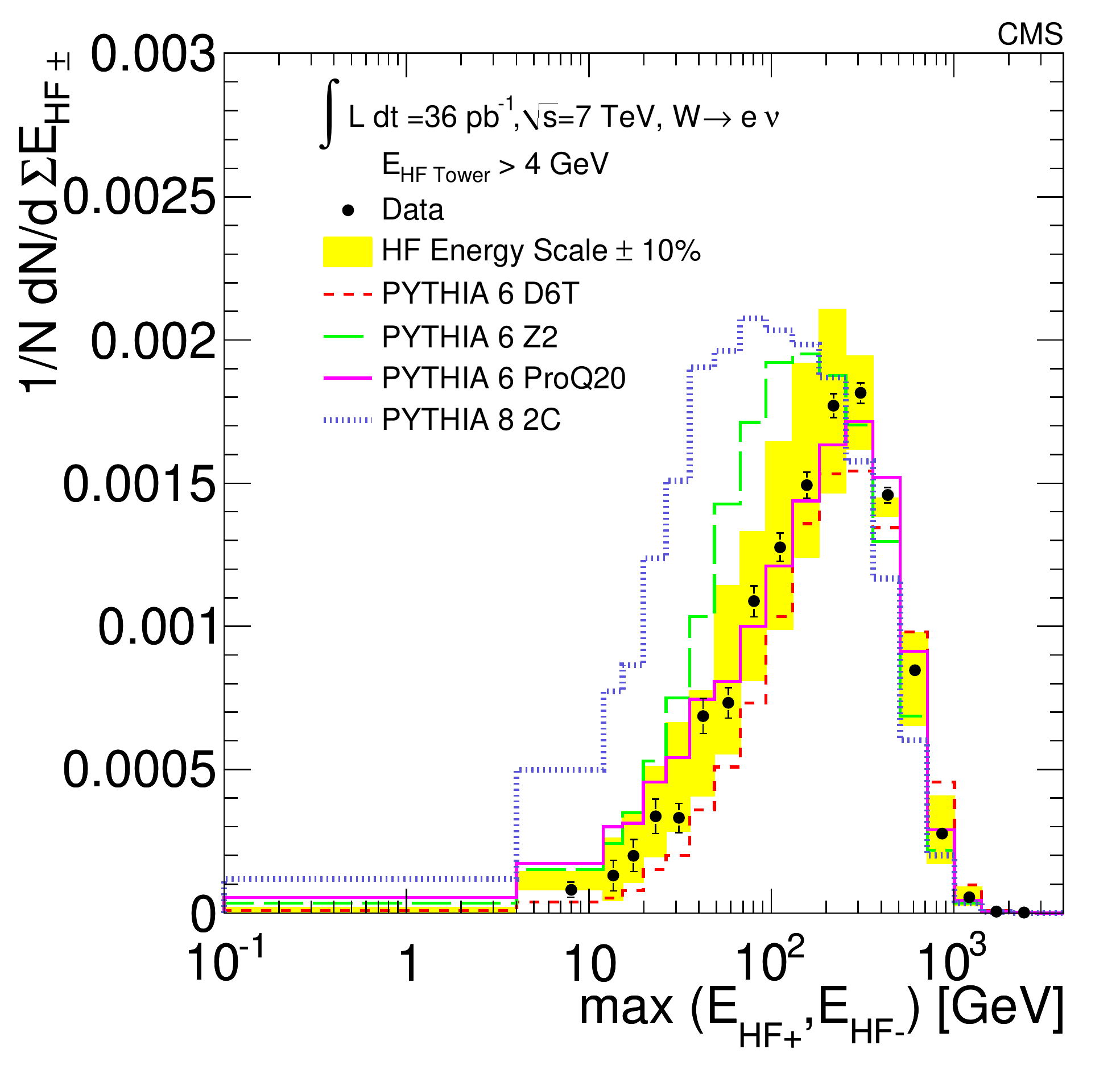}}
\subfigure[]{\label{fig:muMaxEHF}\includegraphics[width=0.5\linewidth]{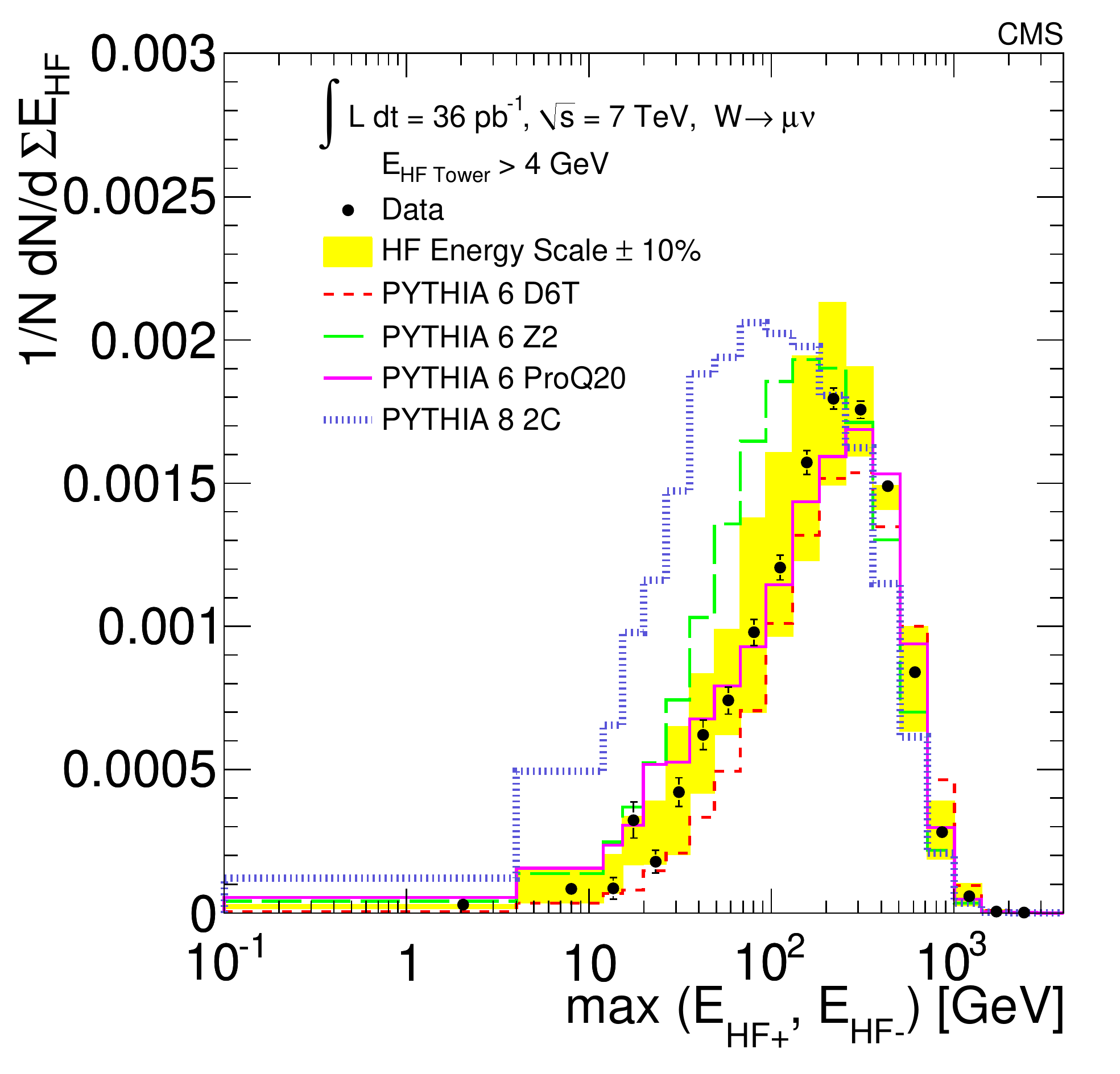} }

\caption{HF energy distributions in $\PW \rightarrow \Pe  \nu X$ (left column) and $\PW \rightarrow \mu  \nu X$ (right column)  events for data and different
MC tunes. The plots (a) and (b)
show the minimum (min ($E_{HF+}, E_{HF-}$)
and (c) and (d) the maximum (max ($E_{HF+}, E_{HF-}$)
of the energy depositions per event in the
HF$+$ and HF$-$ calorimeters.}
\label{Fig5-lowerhigher}
\end{figure*}

\subsection{Interpretation of the observed HF energy and charged-particle multiplicity correlations}
As seen in the previous section, the energy distributions in the two HF calorimeters and
the central charged-particle multiplicities are strongly correlated: large energy depositions in one of the HF calorimeters correspond to large energy depositions in the other HF calorimeter, as well as to an increase in the central charged-particle multiplicity.
Such correlations are also predicted by the
various MC tunes, though with very different strengths. Our observations can be summarized as follows:

{\bf D6T tune:} The inclusive distribution of charged-particle multiplicities, with a minimum $p_{T}$
of 0.5~GeV, is reasonably well described, whereas raising this threshold to $1$~GeV leads to an overestimation of the event rate with large multiplicities. Furthermore, on average much-larger
HF energy depositions are predicted
than observed. When selecting events with small energy depositions in
the HF, the fraction of events in the data is 30-50\% larger than predicted by the D6T tune.
In terms of correlations,
the D6T tune provides a reasonable description only for the charged-particle multiplicity in the medium HF$-$ energy
interval (track $p_{T}$ threshold of $0.5$~GeV) and  for
HF$+$ energy distribution corresponding to the low HF$-$ energy bin.

{\bf Z2 tune:} Overall, the Z2 tune provides a very good description of the inclusive charged-particle multiplicities,
but predicts too many events with very low charged-particle multiplicities. Concerning the HF energy distributions,
too many events with low-energy depositions are predicted.
The correlations between charged-particle multiplicity and HF$-$ energy
are well described. The HF$+$ energy distribution obtained for the low HF$-$ energy interval
is badly modeled, with the MC prediction much higher than the data at low
energies.
However, the correlations are well described for the higher HF$-$ energy intervals.

{\bf Pro-Q20 tune:} This tune provides the best description of the HF energy distributions
and the charged-particle multiplicities with the $p_{T} >  0.5$ GeV
threshold. However, the inclusive charged-particle multiplicity for the $p_{T} >  1.0$ GeV threshold is not well described,
though still closer to the data than the D6T tune.
In terms of correlations, the central charged-particle multiplicities are reasonably well described,
though the fraction of events with large multiplicity and a large HF$-$ energy deposition is
underestimated.
Furthermore, too many events with
low-energy depositions in HF$+$ are predicted when a
low-energy deposition in HF$-$ is selected. For the other HF$-$ energy bins this tune provides a good
description of the data.

{\bf \PYTHIAeight 2C tune:} In the inclusive case, this tune predicts too many events
with low HF energy depositions, whereas the central charged-particle multiplicity distributions are well described.
The HF$+$ energy distributions for the cases of low and medium HF$-$ energy intervals are shifted
towards lower values compared to data, whereas for the high-energy bin good agreement is found.

In summary, none of the analyzed MC tunes provides an overall consistent and reasonable
description of the inclusive charged-particle multiplicities and the HF energy distributions
in the $\PW$ data sample,
as well as correlations between them. It follows that the tunes, which provide a reasonable
description of the underlying event structure for central rapidities in jet events, as presented in
\cite{QCD-10-10},  require substantial modifications to describe the $\PW$ data presented here.
Similar, though statistically less significant results were obtained from the corresponding
$\cPZ$ event samples.

\section{W and Z events with large pesudorapidity gaps}

As the next step, the subset of $\PW$ and $\cPZ$ events with
a single primary vertex and a LRG signature was analyzed.

A LRG event was defined by the requirement
that none of the calorimeter towers had a measured energy of more than 4 GeV in at least one of the HF calorimeters, corresponding to a pseudorapidity interval of 1.9 units.
This subset of events may be enhanced by a diffractive $\PW/\cPZ$ production mechanism.

\subsection{Observed number of LRG events}

Table \ref{Table3-LRGfraction} shows the observed LRG event yields and their
ratio to the number of inclusive $\PW$ and $\cPZ$ single-vertex events
for the three luminosity periods.
This ratio decreases by roughly a factor of 2 to 4 when going from period I to period III.
The decrease can be explained by the HF energy depositions coming from soft pileup events.
As discussed in Section \ref{sec:soft_pile_up} (cf. Fig. \ref{fig:eSumEHF}), adding pileup to the Monte Carlo simulation shifts some of the LRG events to the class of low-energy depositions in the HF.

The inefficiency to detect a vertex in an event with forward energy deposition depends on the
instantaneous luminosity and is estimated from zero-bias events (Section 5.1).
After correcting the observed number of LRG events in the data for pileup effects, using
data,
a constant fraction of LRG events, relative to the total number of $\PW$ and $\cPZ$ events with a single primary vertex, is found for the three instantaneous luminosity periods.
The corrected fraction of LRG events is given in Table \ref{Table4-corrLRGfraction}.
The uncertainties on this correction are small compared to the
statistical errors and to the $\pm$10\% energy scale uncertainties of the HF calorimeters (for details see \cite{pflow}). Indeed, this energy scale variation is the dominant systematic uncertainty for the estimated fraction of LRG events in the data,
resulting in a change of about  $\pm$26\% when varying the tower energy threshold between $3.6$ and
$4.4$ GeV.
{
\begin{table*}[h]
\centering
\caption{Number of LRG events with a single vertex and their percentage relative to all selected $\PW$ and $\cPZ$ events, for the three different luminosity periods and their total.
}

\begin{tabular}{|c|c|c|c|c|}
\hline

 & $ \PW \rightarrow \Pe \nu $  & $ \PW \rightarrow \mu \nu   $ & $ \cPZ\rightarrow \Pe \Pe  $ & $ \cPZ\rightarrow \mu \mu $ \\
\hline
Total   &   100  (0.71\%)        &   145 (0.81\%)  &   19    (0.80\%)    &   23     (0.79\%)  \\
P I      &  17  (1.13\%)         &    31  (1.61 \%)     &   7    (2.7\%)    &     3   (0.91\%)     \\
P II      & 57      (0.72\%)          &   91  (0.86 \%)     &   9  (0.59\%)   &   16  (0.93\%)       \\
P III      & 26     (0.57\%)           &  23  (0.42\%)     &   3  (0.55\%)   &     4  (0.46\%)   \\
\hline
\end{tabular}\vspace{0.3cm}
\label{Table3-LRGfraction}
\end{table*}

{
\begin{table*}[h]
\centering
\caption{Percentage of  LRG events in single-vertex  $\PW$ and $\cPZ$ events, using a pileup correction
determined from data,
for the entire dataset and the three different luminosity periods.
Only the statistical uncertainties are given; the dominant systematic uncertainty from the
HF energy scale is about $\pm26\%$.
}
\begin{tabular}{|c|c|c|c|c|}
\hline
 & $ \PW \rightarrow \Pe \nu $  & $ \PW \rightarrow \mu \nu   $ & $ \cPZ\rightarrow \Pe \Pe  $ & $ \cPZ\rightarrow \mu \mu $ \\
\hline
Total     & 1.37 $\pm$ 0.14\%            & 1.50 $\pm$ 0.13 \%       &   1.73 $\pm$ 0.43\%     & 1.49 $\pm$ 0.31\%  \\
P I       &    1.68 $\pm$  0.41  \%      & 2.39 $\pm$  0.43 \%      &    5.52$\pm$  2.08\%    &  1.36 $\pm$ 0.78 \%       \\
P II      &   1.27 $\pm$ 0.17 \%         & 1.54  $\pm$ 0.16 \%      &  1.57 $\pm$  0.52\%     &  1.65 $\pm$ 0.41\%        \\
P III     &  1.53 $\pm$  0.30  \%        & 1.12 $\pm$ 0.23 \%       &   1.47$\pm$  0.85\%     & 1.22 $\pm$ 0.61\%         \\
\hline
\end{tabular}\vspace{0.3cm}

\label{Table4-corrLRGfraction}

\end{table*}

Combining the results obtained with electrons and muons, the percentage of $\PW$ and $\cPZ$ events
with LRG signature is
($1.46 \pm 0.09$ (stat.) $\pm$ 0.38 (syst.))\% and ($1.57 \pm 0.25$ (stat.) $\pm$ 0.42 (syst.))\%,
respectively.
In comparison, as can be seen from Figs. \ref{Fig2-mult} and \ref{Fig5-lowerhigher},
the fraction of $\PW$ LRG events predicted with the \PYTHIAsix Z2 and Pro-Q20
and the \PYTHIAeight 2C tunes are larger than observed in the data.
In contrast, for the D6T tune the number of LRG events is smaller than in the data.

\subsection{Jet activity in W/Z events with a LRG signature and search for exclusive W/Z production}

A further subset of $\PW$ and $\cPZ$ events are those that show some jet activity,
using the particle-flow algorithm with a cone size of 0.5. We find
$(11.1 \pm 0.2)$\% of the selected  $\PW$ and $\cPZ$ events with a single vertex contain at least one reconstructed jet
with a transverse momentum above 30~GeV and $|\eta| < 2.5$. Taking the subsample
of 100 identified $\PW \rightarrow \Pe \nu$ events with a LRG signature, 8 events are found
with one or more jets above a 30 GeV threshold. The corresponding numbers for the muon channel are 145 identified LRG events and 8 of them with at least one jet.
Thus, we find that (6.5$\pm$1.6)\% of the LRG events have jet activity, which
is smaller (but still consistent) with the fraction of events with jets
observed in the inclusive $\PW$ sample. No other particular features of the events with jet activities, when compared to the MC simulations, are observed.

Another potentially interesting class of LRG events consists of $\PW$ and $\cPZ$
events with essentially no activity besides that from the vector-boson decays.
Such candidate exclusive events are selected with the requirement that both HF calorimeters fulfill the
LRG condition and that no particle-flow object besides the lepton(s)
is reconstructed in the central detector, above a transverse momentum threshold of $0.5$~GeV.
For the electron selection no such events are found.
In the muon case, 2 $\PW$ and 2 $\cPZ$ event candidates with zero energy in both
HF calorimeters are found. All four events have some reconstructed tracks in the central detector.
The number of observed events is consistent with expected number of non-exclusive
$\PW$ and $\cPZ$ events predicted from the MC simulations.

\subsection{Size of the pseudorapidity gap and central gaps}

For the study of events with large pseudorapidity gaps an interesting parameter is how far the
size of the gap
extends into the central detector. One might intuitively
expect that the signal for diffractive events, compared to the effects from
multiplicity fluctuations, would become stronger when the
the gap size increases into the central detector.  This intuitive view is confirmed when comparing
diffractive $\PW$ events simulated with \POMPYT , where the decrease in event yields with increasing gap size is much smaller than in the different non-diffractive MC models.

Obviously, the definition of the gap is ambiguous, as the meaning of
zero activity or zero energy depositions depends on the experimental criteria for the detection of
particles in the data and in the MC simulation. For this study the size of the pseudorapidity gap was determined
by using particle-flow objects with a minimum energy of $1.5$ GeV for $|\eta| < 1.5$ (barrel),  $2$ GeV for $1.5 < |\eta| < 2.85 $ (endcaps),
and $4$ GeV for $|\eta| > 2.85$ (HF calorimeters). For charged particle-flow candidates a minimum transverse momentum of $0.5$ GeV was required. The largest ($\eta_{max}$) and smallest ($\eta_{min}$) observed pseudorapidity values of the particles are used to determine the gap size between the
maximum (minimum) $\eta$ coverage of the experiment and the nearest detected particle on each side.
In order to combine both hemispheres, we
define $\tilde{\eta}$ as the minimum of $\eta_{max}$ and  $-\eta_{min}$.
The size of the pseudorapidity gap is then $\Delta \eta_{gap}^{4.9} = 4.9 - \tilde{\eta}$, where 4.9 is the largest $\eta$ value covered by the HF.

Figures \ref{fig:eEtaMax} and \ref{fig:muEtaMax} show the $\tilde{\eta}$ distribution in the data and MC simulation
with different tunes,  for the $\PW$ decays to electrons and muons, respectively.
The fraction of events with pseudorapidity gaps decreases rapidly when the gap size increases.
A statistically significant excess of events at $\tilde{\eta}$ values smaller than 3 is observed,
compared to the predictions of the non-diffractive model implemented in \PYTHIAsix, tune D6T.
However, the fraction of events with very large gaps and without a diffractive component
in the \PYTHIAsix Z2, Pro-Q20, and \PYTHIAeight 2C tunes and up to the largest observed gap size
is larger than in the data.

The stability of the $\tilde{\eta}$ distribution was tested by
allowing a $\pm$10\% variation of the particle-flow candidate energy and momentum thresholds in the data. The resulting variations were found to be similar to the statistical uncertainties.

If $\tilde{\eta} < 0$, all the reconstructed particle-flow objects in the event are contained in one hemisphere.
Combining the $\PW$ events with LRG signature in both lepton channels,
4 events with one ``empty'' detector hemisphere, corresponding to a gap of at
least $\Delta \eta_{gap}^{4.9} = 4.9$ units in pseudorapidity, are observed.
In comparison, 0.8, 3.5, and 2.2 such events are expected from the non-diffractive MC simulation
based on the \PYTHIAsix D6T, Z2, and Pro-Q20 tunes, respectively.

As discussed in Sections \ref{sec:eventselection} and
\ref{sec:trackmult_energyflow}, soft pileup events without a detectable second vertex
remain in the sample. However, since these events do not produce significant particle-flow
in the central pseudorapidity regions, the effect on events with pseudorapidity gaps in the more central
region, $ |\eta| < 2.85 $, is expected to be small.  The number of pseudorapidity gap events
in this detector region, when ignoring the information in the HF detectors,
should mainly depend on the amount of  very low multi-parton activity and thus on
the number of low-multiplicity events.
The $\tilde{\eta}$ distributions using only particle-flow objects with $|\eta| < 2.85$ and
ignoring the information in the HF are shown for $\PW$ events in Figs. \ref{fig:eEtaMaxNoHF} and \ref{fig:muEtaMaxNoHF}. Accordingly, the gap size is now defined as $\Delta \eta_{gap}^{2.85} = 2.85 - \tilde{\eta}$. Again, when compared to the MC simulation with the D6T tune,
the data show a large excess of events with $\tilde{\eta}$ below 1,
corresponding to a central pseudorapidity gap of $\Delta \eta_{gap}^{2.85} \geq 1.85$. The fraction
of such gap events in the data is reasonably well described by the
\PYTHIAsix Z2 and Pro-Q20 tunes, and much larger fractions are predicted by the
\PYTHIAeight 2C tune.

The limited number of LRG events,
as well as the large uncertainties related to the modeling of the underlying event and multi-parton interactions,
prevent any conclusions from being drawn on the possible presence of a diffractive
$\PW/\cPZ$-production component from the observed rate of events with a pseudorapidity gap in the
central detector.

\begin{figure*}[hbtp]
\subfigure[]{\label{fig:eEtaMax}\includegraphics[width=0.5\linewidth]{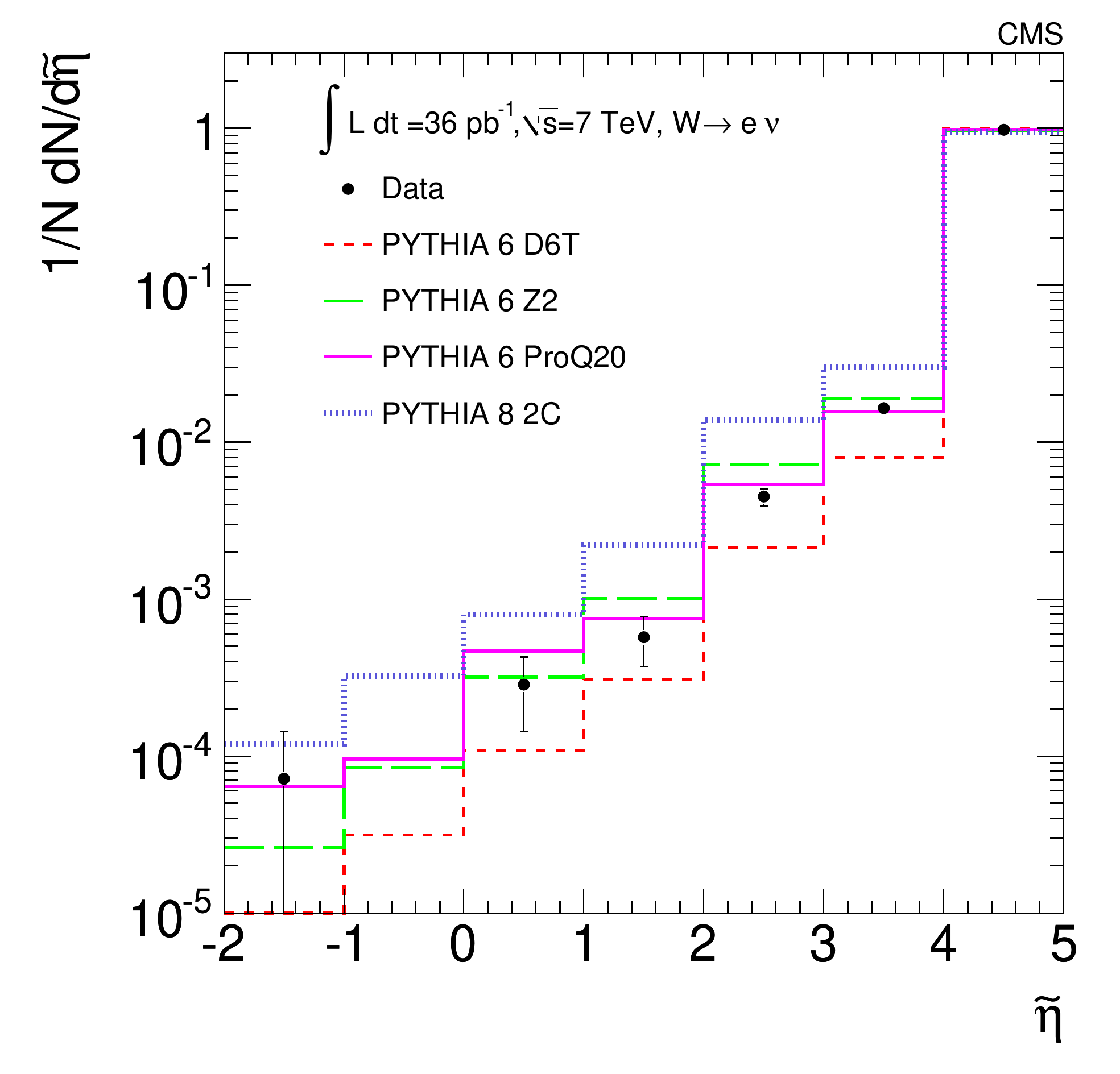}}
\subfigure[]{\label{fig:muEtaMax}\includegraphics[width=0.5\linewidth]{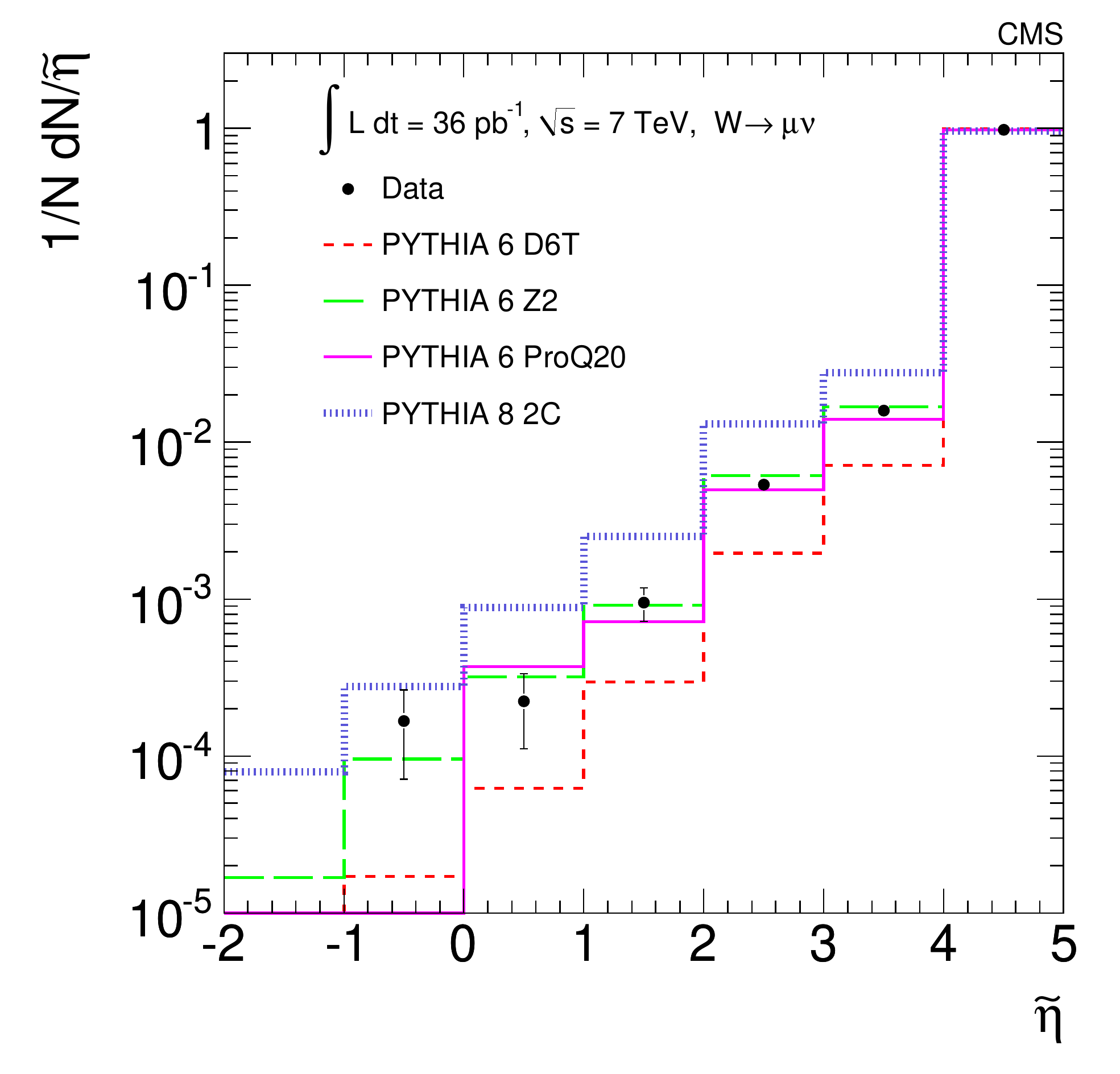}}  \\
\subfigure[]{\label{fig:eEtaMaxNoHF}\includegraphics[width=0.5\linewidth]{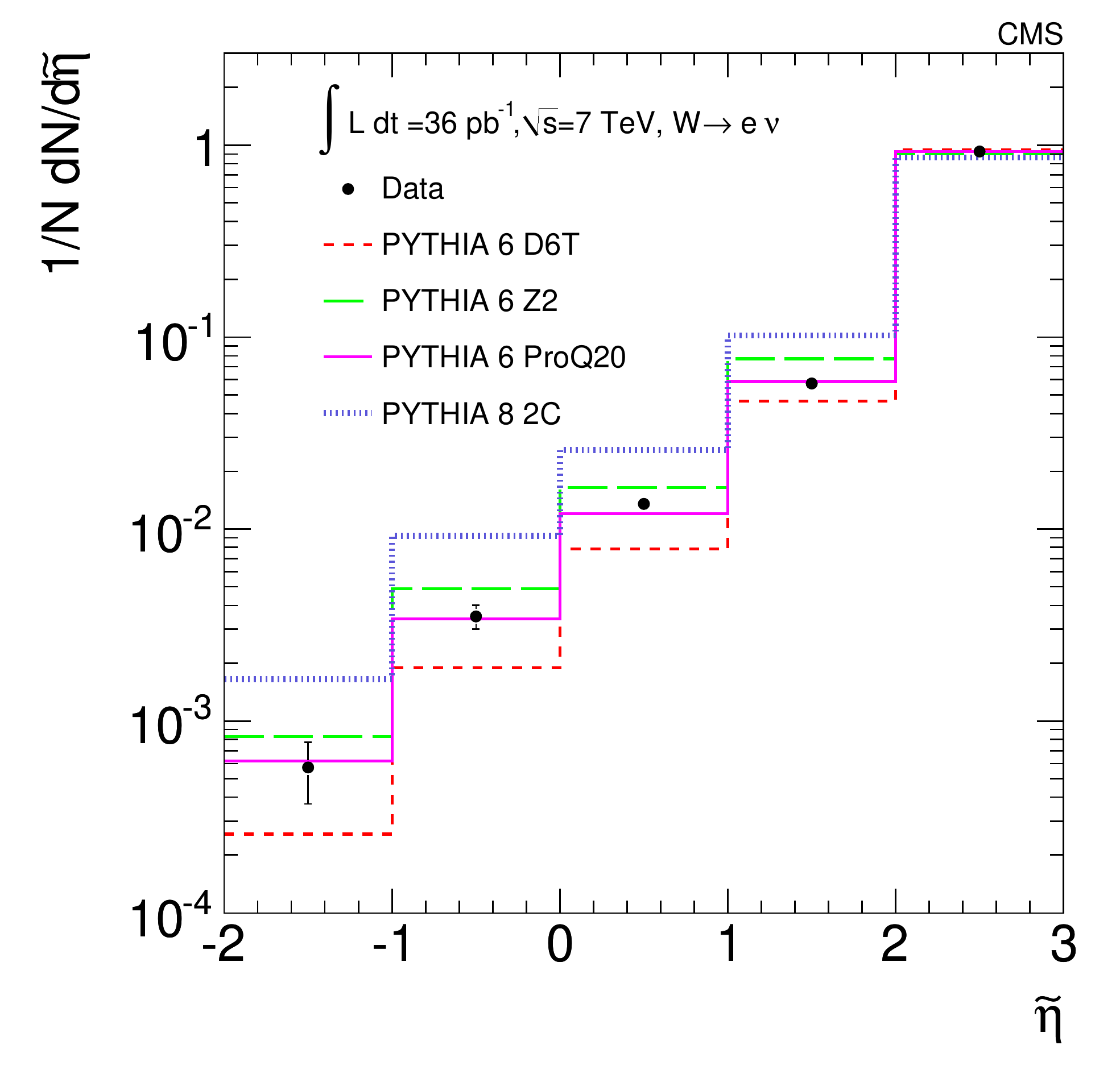}}
\subfigure[]{\label{fig:muEtaMaxNoHF}\includegraphics[width=0.5\linewidth]{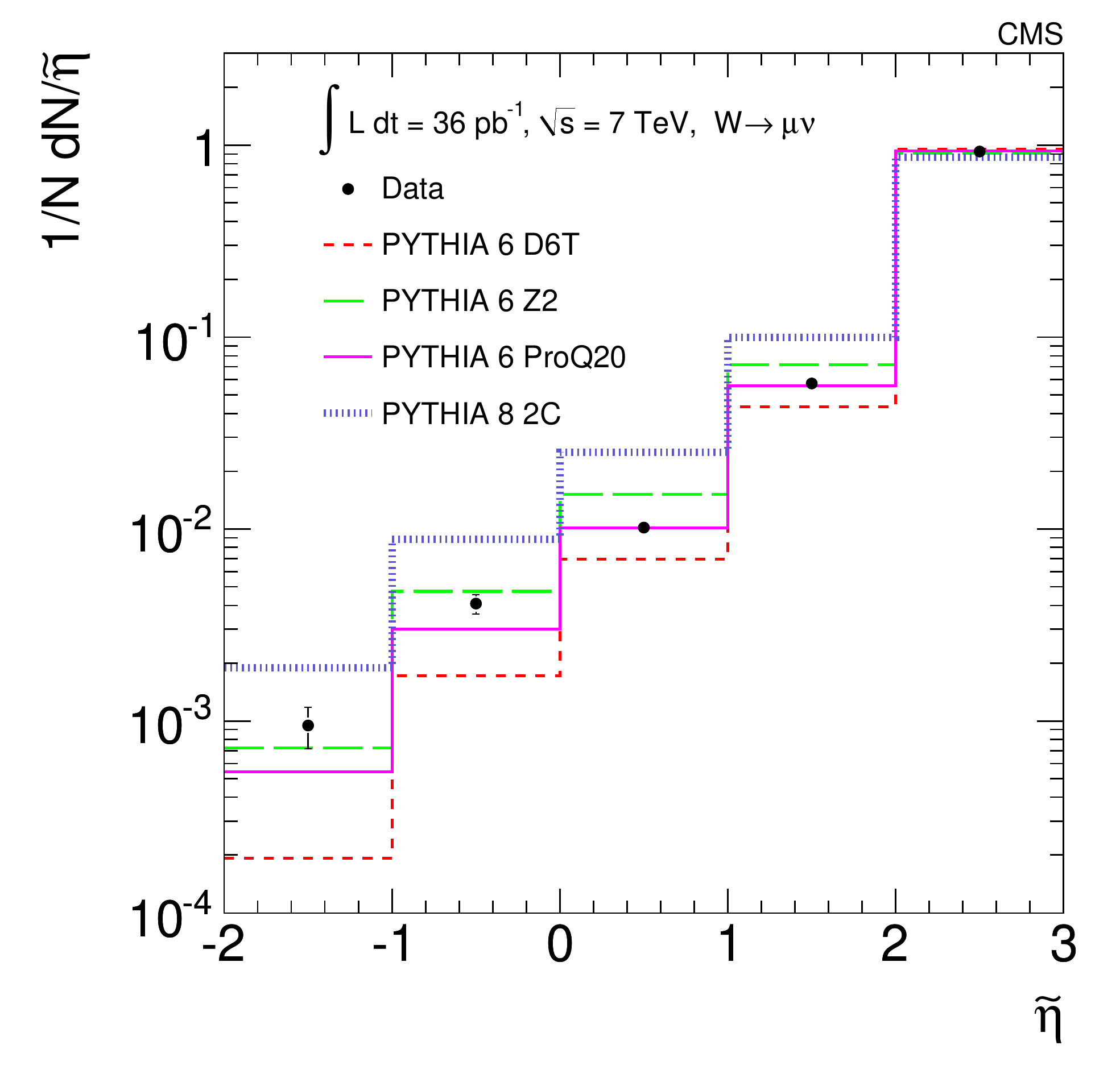}}

\caption{The $\tilde{\eta}$ distribution for $\PW$ events with (a) electron
and (b) muon
decays in data and for various MC simulations.
The corresponding distributions ignoring the
HF calorimeter information are shown in (c) and (d).
}
\label{Fig6-etamax}
\end{figure*}

\subsection{Charged-particle multiplicity and forward energy distributions in LRG events}

The charged-particle multiplicity distribution in LRG events,
from combining the $\PW$ events in the electron and muon channels, is shown in Fig. \ref{fig:emuTracksLRG}
for a minimum track $p_{T}$ of $0.5$~GeV.
A slight excess of events with large charged-particle multiplicities is found in the data,
compared to the various MC tunes. However, overall the track $p_{T}$ spectrum
is well described. The number of LRG events with 20 and more tracks, combining the electron and muon channels, is 33 in the data. Only 13 (19) events with more than 20 tracks
are expected from the D6T (Z2) tunes. A similar, but statistically less significant, excess of
events with multiplicities larger than predicted by the different tunes
is also observed when a track threshold of $p_{T} > 1.0$~GeV is required.

The \POMPYT diffractive model, which does not include multi-parton interactions,
predicts even smaller charged-particle multiplicities. However, the observed
excess of events with relatively large charged-particle multiplicities in LRG events
could be an indication of a diffractive component in the multi-parton interactions,
as depicted in Figs. \ref{fig:diag-c} and \ref{fig:diag-e}.

The corresponding distribution for the energy sum in the HF calorimeter opposite to the gap is shown in
Fig. \ref{fig:emuSumEHFLRG}. The average total energy of 150 GeV with an r.m.s. of 160 GeV
is about a factor of two smaller than the one observed for the inclusive HF energy distribution,
and is reasonably well
described by the various MC tunes. In the data, we find 2 events with no towers above
the energy threshold of 4 GeV in either HF calorimeters,  in agreement with the expectation from
the D6T tune. This number is slightly lower, but still consistent,
with the expections from the Z2 and the \PYTHIAeight 2C tunes. The Pro-Q20 tune predicts 8 such events.

\begin{figure*}[htp]
\subfigure[]{\label{fig:emuTracksLRG}\includegraphics[width=0.5\linewidth]{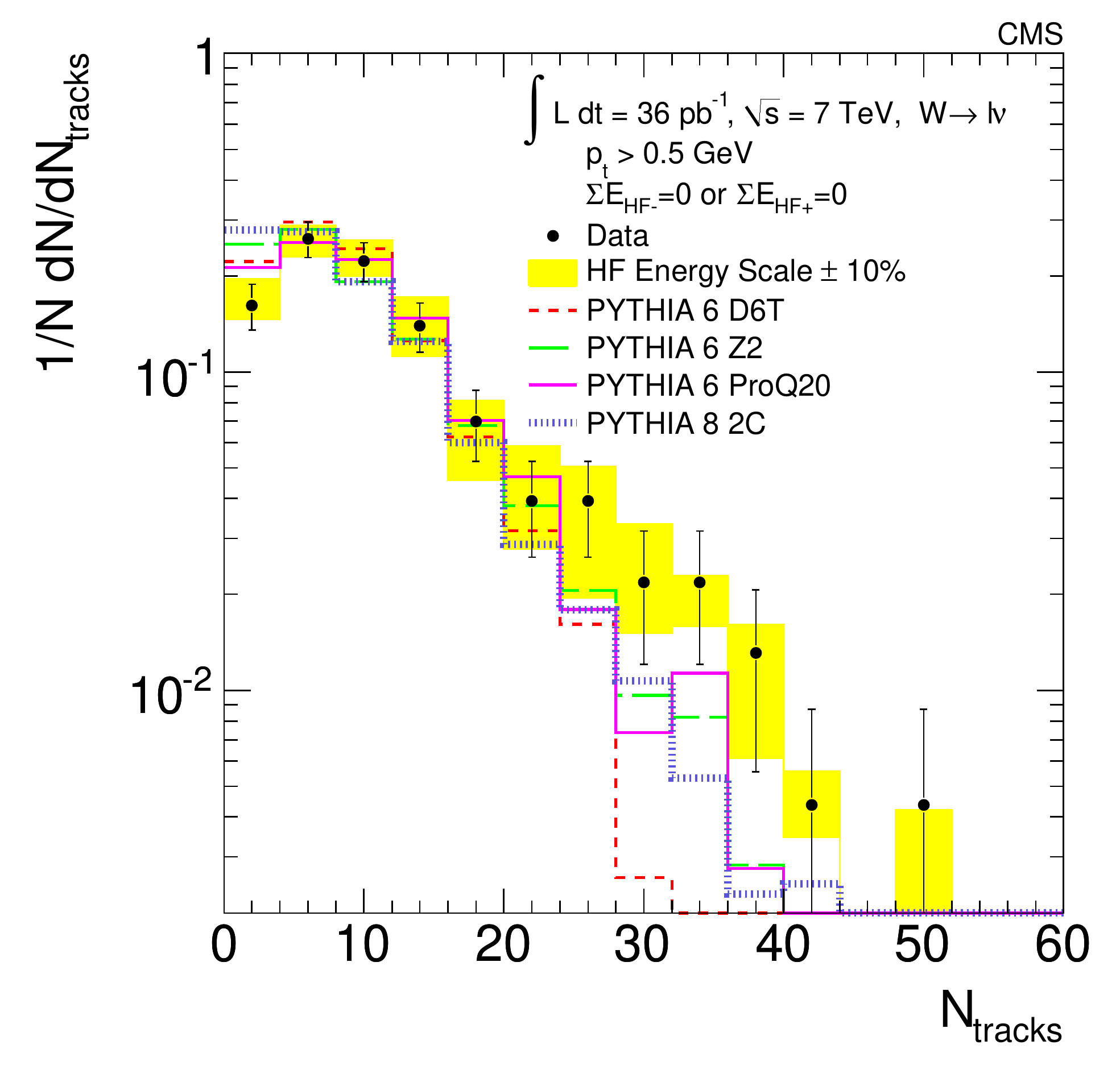}}
\subfigure[]{\label{fig:emuSumEHFLRG}\includegraphics[width=0.5\linewidth]{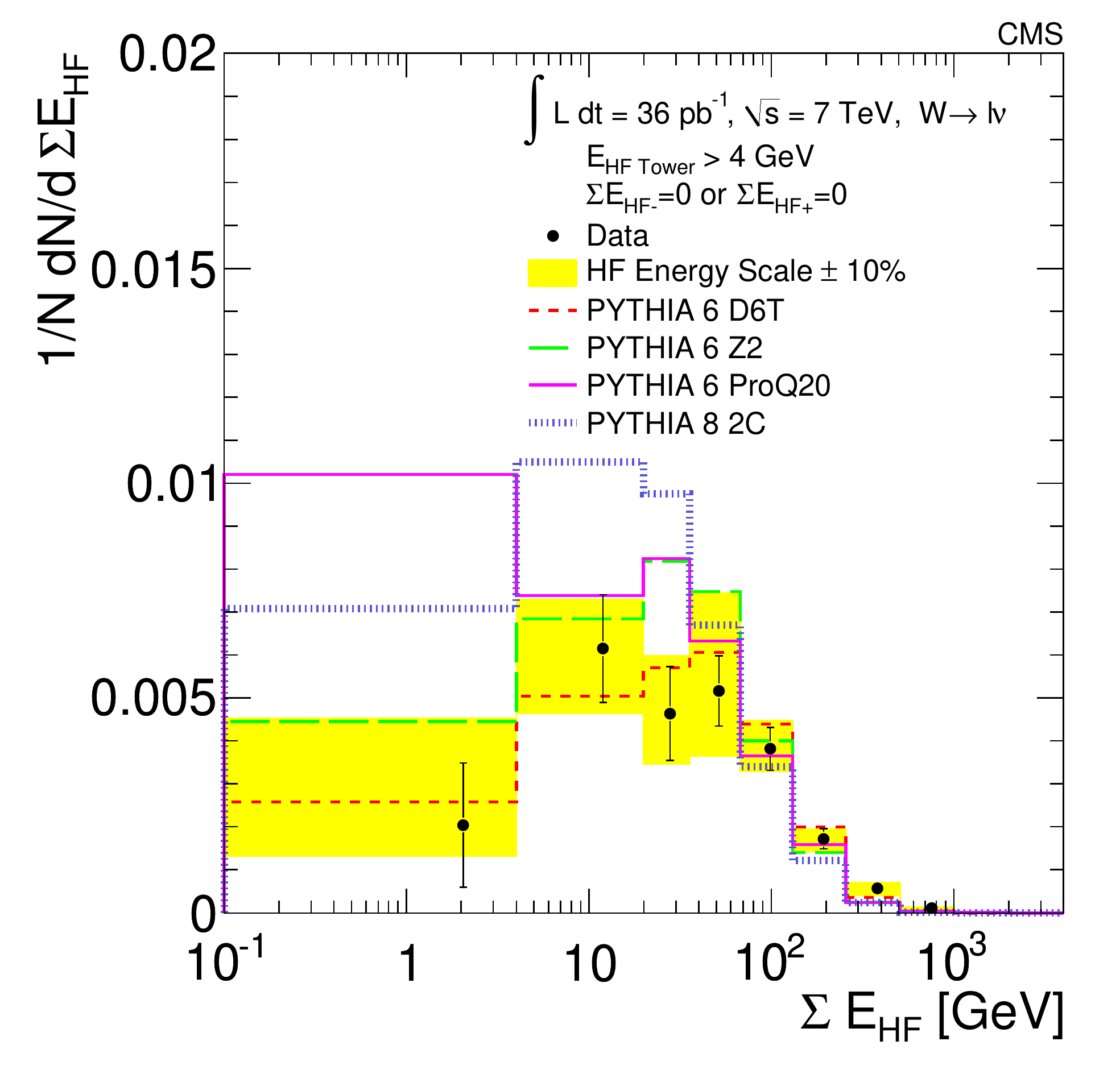}}

\caption{(a) Charged-particle multiplicity
and (b) HF energy
distributions (opposite to the gap) in $\Pp\Pp \rightarrow \PW^\pm X \rightarrow \ell^{\pm} \nu X$
events with a LRG signature, for the data and different MC tunes.
The charged-particle multiplicity distribution is obtained for a track $p_{T}$ threshold  of  $0.5$ GeV.}
\label{combinedTracksLRG}
 \end{figure*}

\subsection{Hemisphere correlations between the gap and the \texorpdfstring{$\mathbf{W}$ ($\mathbf{Z}$)}{W (Z)} boson}

Figure \ref{fig:combinedCorrelation} shows the distribution of the signed charged lepton pseudorapidity $\eta_{\ell}$
in $\PW$ events with a LRG signature (electron and muon channels combined). The sign is defined
to be positive when the gap and the lepton are in the same hemisphere and negative
otherwise. The data show that charged leptons
from $\PW$ decays are found more often in the hemisphere opposite to the gap.
Combining the electron and muon channels, 147 events are found with the charged lepton
in the hemisphere opposite to the gap and 96 events with the lepton in the same hemisphere.
Defining an asymmetry as the ratio of the difference between the numbers of LRG events
in each hemisphere and the sum, the corresponding asymmetry is (-21.0 $\pm$  6.4)\%.
In the case of $\cPZ$ candidates (the rapidity of the $\cPZ$ is used) with a LRG signature,
24 (16) events are in the opposite (same) hemisphere as the gap,
resulting in an asymmetry of (-20 $\pm$ 16)\%.

In comparison, the various non-diffractive MC tunes predict a symmetric lepton
pseudorapidity distribution in LRG events. On the other hand,
events generated with the \POMPYT generator, based on a diffractive production model,
exhibit a strong asymmetry.
This can be explained in terms of diffractive PDFs, which peak at smaller $x$ (the parton momentum
divided by the proton momentum) than the conventional
proton PDFs. The produced $\PW(\cPZ)$ is thus boosted in the direction of the parton that had the
larger $x$. This is typically the direction of the dissociated proton, i.e., opposite to the gap.
The signed lepton pseudorapidity distribution in the data is fit
to the predictions from the diffractive \POMPYT and the non-diffractive
\PYTHIA event generators, with the relative fraction of the two
as the free parameter.
The fit results in a fraction of diffractive events in the LRG sample of
(50.0 $\pm$ 9.3 (stat.) $\pm$ 5.2 (syst.))\%,
assuming the model of diffraction
implemented in \POMPYT and using the
\PYTHIAsix Pro-Q20 for the simulation of non-diffractive events.
The fit results are shown in Fig.\ \ref{fig:combinedCorrelation}.
The fits using the combination of \POMPYT with other tunes give similar results, and
only the non-diffractive contribution from the other tunes is shown in Fig. \ref{fig:combinedCorrelation}.
The systematic uncertainty of 5.2\% has been determined from the 10\% HF energy scale variations
and from the fits with the different tunes, using the maximal and minimal fractions obtained from the different fits.

\begin{figure}[htp]
\includegraphics[width=\columnwidth]{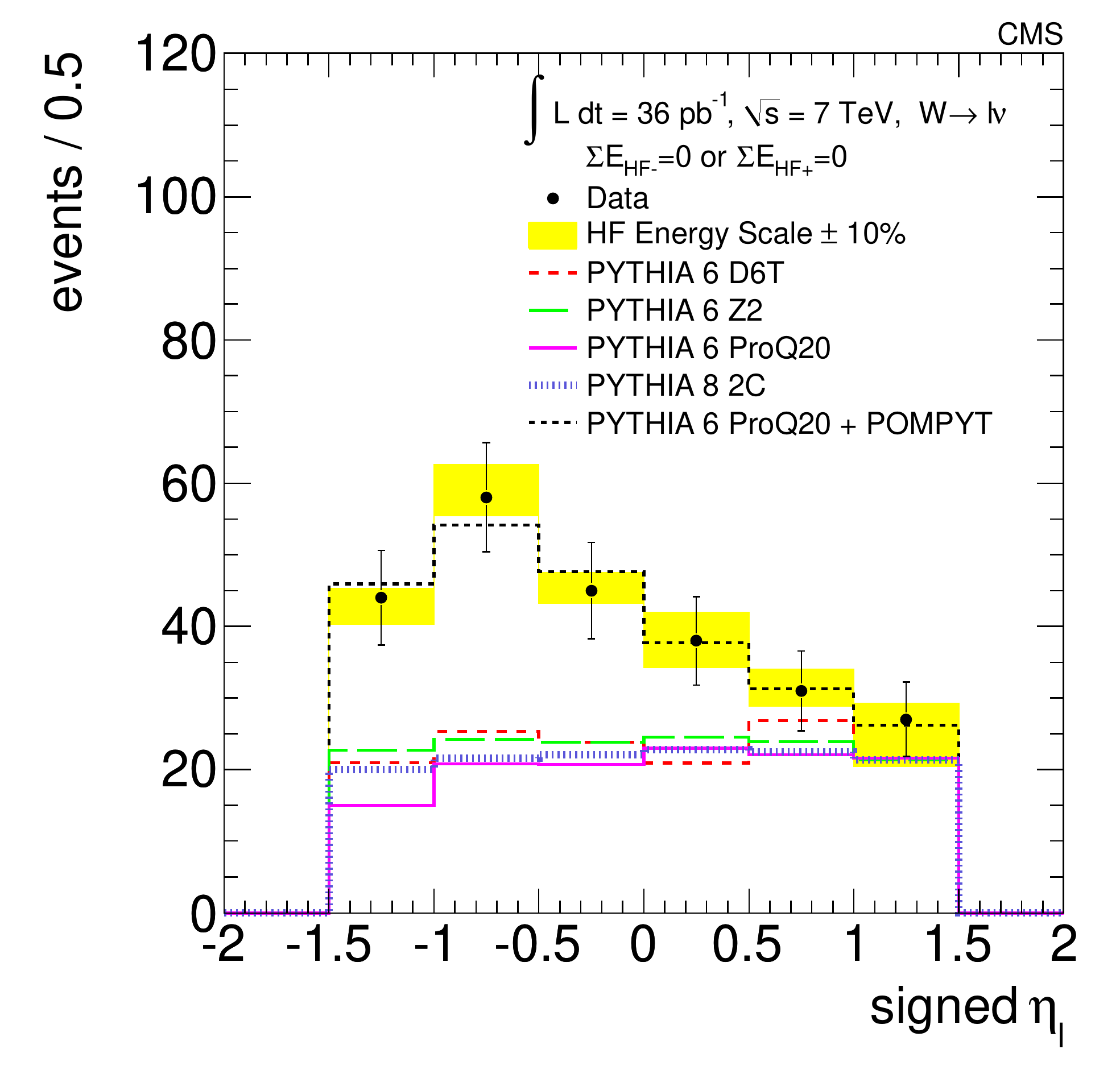}

  \caption{Signed lepton pseudorapidity distribution in $\PW$ events with a LRG signature,
with the sign defined by the pseudorapidity of the lepton relative to the
gap (positive for the lepton and gap in the same hemisphere, negative otherwise). Electron and muon channels are combined.
The fit result for the combination of \PYTHIAsix (Pro-Q20 tune) and \POMPYT predictions
is shown as a dotted black line. For the other \PYTHIA tunes, only the non-diffractive component is shown.
}
 \label{fig:combinedCorrelation}
 \end{figure}

The asymmetry in the signed $\eta_{\ell}$ distribution for non-LRG events decreases
when the forward energy deposition increases. For example, for HF energy depositions
in the intervals 20-100~GeV, 200-400~GeV, and $> 500$~GeV, the
asymmetry is (-3.5 $\pm$ 1.1)\%, (-2.7 $\pm$ 1.0)\%, and (0.9 $\pm$ 2.3)\%.
The small residual asymmetry in events with low HF energy depositions is insignificant in comparison to the one in LRG events. However, this could be explained by the presence of a small fraction of diffractively produced $\PW$ bosons events in which the
LRG signature is destroyed by the accompany multi-parton interaction or by the undetected pileup component.
For higher energy depositions in the forward region, the asymmetry vanishes.

\section{Conclusions}

Central charged-particle multiplicities, forward energy flow, and correlations between them have been
studied in $\PW$ and $\cPZ$ events, identified by the vector-boson decays to electrons and
muons, using the 2010 data sample of $\Pp\Pp$ collisions at 7 TeV, corresponding to an integrated luminosity of 36 pb$^{-1}$.

None of the studied MC tunes provides simultaneously a satisfactory description of the
charged-particle multiplicity in the
central pseudorapidity region ($|\eta| < 2.5$) and the
forward energy flow ($3 < |\eta| < 4.9$).
The \PYTHIAsix Z2 and \PYTHIAeight 2C tunes give a reasonable description of the central
charged-particle multiplicity, but predict too many events with relatively
low-energy depositions in the forward calorimeters.
The \PYTHIAsix D6T tune predicts too many events with high charged-particle multiplicities,
too few events with low-energy depositions, and
too many events with very large energy depositions in the forward calorimeters.
The Pro-Q20 tune provides the best description of the forward energy distribution
and a good description of the charged-particle multiplicity, when a track $p_{T}$ threshold of $0.5$ GeV is applied.
However, the charged-particle multiplicity with $p_{T} >  1.0$~GeV is not well described,
though the prediction is closer to the data than that for the D6T tune.

Strong positive correlations between the energy measured in the two forward calorimeters
(i.e. at positive and negative rapidities) and the charged-particle multiplicity are observed in the data and in Monte Carlo models.
However, the correlations in the various MC tunes are
different from those seen in the data.

As far as the LRG events are concerned,
the following observations can be made:
\begin{itemize}
\item Out of a sample of about 40 000 $\PW$ and $\cPZ$ events, almost 300 events with a LRG signature
are found.
According to the \POMPYT model of diffractive $\PW$ production, such events can
be interpreted as diffractive. However, while
the observed fraction of such events is significantly larger than predicted with the non-diffractive
\PYTHIAsix D6T tune, it is  smaller than expected from the \PYTHIAsix Z2, Pro-Q20,
and \PYTHIAeight 2C tunes. Thus, no conclusions can be drawn from the presence of LRG events.
\item The central charged-particle multiplicity in these events is somewhat larger
than predicted by the various models.
\item The HF energy distribution opposite to the gap peaks at much smaller values
than in the inclusive events, and is reasonably well described by the MC generators.
\item A large asymmetry
is observed between the number of events with
the charged lepton (from the $\PW$ decay) in the opposite and with it in the same hemisphere as
the pseudorapidity gap.
Such an asymmetry is predicted by \POMPYT,  in contrast to the various non-diffractive
\PYTHIA MC tunes. When fitting the observed asymmetry in LRG events with an admixture of
diffractive (\POMPYT) and non-diffractive (\PYTHIA) events,
the diffractive component is determined to be
(50.0 $\pm$ 9.3 (stat.) $\pm$ 5.2 (syst.))\%, thus providing the first evidence for diffractive $\PW$ production
at the LHC. A comparable, but statistically less significant asymmetry is seen in $\cPZ$ events with large pseudorapidity gaps.
\end{itemize}

\section*{Acknowledgments}

We wish to congratulate our colleagues in the CERN accelerator departments for the excellent
performance of the LHC machine. We would like to thank P. Skands for many explanations and discussions
concerning the different underlying event tunes.
We would also like to thank the technical and administrative staff at CERN and
other CMS institutes, and acknowledge support from: FMSR (Austria); FNRS and FWO (Belgium);
CNPq, CAPES, FAPERJ, and FAPESP (Brazil); MES (Bulgaria); CERN; CAS, MoST, and NSFC (China);
COLCIENCIAS (Colombia); MSES (Croatia); RPF (Cyprus); Academy of Sciences and NICPB
(Estonia); Academy of Finland, ME, and HIP (Finland); CEA and CNRS/IN2P3 (France); BMBF, DFG,
and HGF (Germany); GSRT (Greece); OTKA and NKTH (Hungary); DAE and DST (India); IPM (Iran);
SFI (Ireland); INFN (Italy); NRF and WCU (Korea); LAS (Lithuania); CINVESTAV, CONACYT, SEP,
and UASLP-FAI (Mexico); PAEC (Pakistan); SCSR (Poland); FCT (Portugal); JINR (Armenia,
Belarus, Georgia, Ukraine, Uzbekistan); MST and MAE (Russia); MSTDS (Serbia); MICINN and CPAN
(Spain); Swiss Funding Agencies (Switzerland); NSC (Taipei); TUBITAK and TAEK (Turkey); STFC
(United Kingdom); DOE and NSF (USA). Individuals have received support from the Marie-Curie
IEF program (European Union); the Leventis Foundation; the A. P. Sloan Foundation; the
Alexander von Humboldt Foundation; and the Associazione per lo Sviluppo Scientifico e
Tecnologico del Piemonte (Italy).

\bibliography{auto_generated}   % will be created by the tdr script.
\cleardoublepage \appendix\section{The CMS Collaboration \label{app:collab}}\begin{sloppypar}\hyphenpenalty=5000\widowpenalty=500\clubpenalty=5000\input{FWD-10-008-authorlist.tex}\end{sloppypar}
\end{document}

%% file: FWD-10-008-authorlist.tex
\textbf{Yerevan Physics Institute,  Yerevan,  Armenia}\\*[0pt]
S.~Chatrchyan, V.~Khachatryan, A.M.~Sirunyan, A.~Tumasyan
\vskip\cmsinstskip
\textbf{Institut f\"{u}r Hochenergiephysik der OeAW,  Wien,  Austria}\\*[0pt]
W.~Adam, T.~Bergauer, M.~Dragicevic, J.~Er\"{o}, C.~Fabjan, M.~Friedl, R.~Fr\"{u}hwirth, V.M.~Ghete, J.~Hammer\cmsAuthorMark{1}, S.~H\"{a}nsel, M.~Hoch, N.~H\"{o}rmann, J.~Hrubec, M.~Jeitler, W.~Kiesenhofer, M.~Krammer, D.~Liko, I.~Mikulec, M.~Pernicka, B.~Rahbaran, H.~Rohringer, R.~Sch\"{o}fbeck, J.~Strauss, A.~Taurok, F.~Teischinger, C.~Trauner, P.~Wagner, W.~Waltenberger, G.~Walzel, E.~Widl, C.-E.~Wulz
% ...
\vskip\cmsinstskip
% ...
1:~~Also at CERN, European Organization for Nuclear Research, Geneva, Switzerland\\
% ...
